\documentclass[%
aps,
 amsmath,amssymb,
]{revtex4-2}

\usepackage{graphicx}
\usepackage{dcolumn}
\usepackage{bm}

\usepackage[utf8]{inputenc}
\usepackage{float}

\usepackage{enumitem}
\usepackage[english]{babel}
\usepackage{amssymb}
\usepackage{amsmath,mathtools}
\usepackage{xcolor}
\usepackage{graphicx}
\usepackage{multirow}
\usepackage{overpic}
\usepackage{psfrag}
\usepackage{bm}
\usepackage{url}
\usepackage{epsfig}
\usepackage{marvosym}
\usepackage{verbatim}
\usepackage{siunitx}
\usepackage{rotating}
\usepackage{subcaption}
\usepackage{caption}
\usepackage{hyperref}
\hypersetup{colorlinks=true,linkcolor=red,citecolor=blue}
\usepackage{ulem}
\usepackage{lmodern}
\usepackage{booktabs} 
\usepackage{moresize}

\usepackage{lineno}
\usepackage{lipsum}
\usepackage{tikz}
\usepackage{pgfplots}
\usepgfplotslibrary{colormaps} 
\usetikzlibrary{pgfplots.colormaps} 
\usepgfplotslibrary{colorbrewer}
\pgfplotsset{compat=1.14}
\usepgfplotslibrary{fillbetween}
\usepgfplotslibrary{statistics}
\usepgfplotslibrary{groupplots} 
\usetikzlibrary{shapes, arrows, arrows.meta, calc, positioning, decorations,
decorations.pathmorphing, decorations.text, decorations.markings, fit, plotmarks}
\pgfplotsset{translate gnuplot=true}
\usepackage{etex} 								

\definecolor{myblue1}	{RGB}{0,177,234}				
\definecolor{myblue2}	{RGB}{76,200,239}				
\definecolor{myblue3}	{RGB}{127,215,244}				
\definecolor{myblue4}	{RGB}{178,231,248}				
\definecolor{myblue5}	{RGB}{198,251,255}				
\definecolor{mybluegray1}{RGB}{0,127,167}				
\definecolor{mybluegray2}{RGB}{76,165,193}				
\definecolor{mybluegray3}{RGB}{127,191,211}				
\definecolor{mybluegray4}{RGB}{178,216,228}				
\definecolor{mygray1}	{RGB}{76,84,93}				
\definecolor{mygray2}	{RGB}{129,135,141}				
\definecolor{mygray3}	{RGB}{165,169,174}				
\definecolor{mygray4}	{RGB}{201,203,206}				
\definecolor{myorange1}	{RGB}{255,126,46}				
\definecolor{myorange2}	{RGB}{255,164,108}				
\definecolor{myorange3}	{RGB}{255,190,150}				
\definecolor{myorange4}	{RGB}{255,216,192}				
\definecolor{mypurple1}{RGB}{89,89,171}
\definecolor{mypurple4}{RGB}{189,189,231}

\pgfplotsset{
    colormap={custom}{
        rgb255=(54,54,255)
        rgb255=(255,255,255)
        rgb255=(255,6,6)
    }}

\pgfplotsset{%
  compat=newest,
  colormap={redyellow}{
  color=(red) color=(yellow)
  },
}

\pgfplotsset{
/pgfplots/colormap={hot2}{[1cm]rgb255(0cm)=(0,0,0) rgb255(3cm)=(255,0,0)
rgb255(6cm)=(255,255,0) rgb255(8cm)=(255,255,255)}
}

\pgfplotsset{
/pgfplots/colormap={temp}{[1cm] rgb255=(36,0,217) rgb255=(25,29,247) rgb255=(41,87,255)
rgb255=(61,135,255) rgb255=(87,176,255) rgb255=(117,211,255) rgb255=(153,235,255)
rgb255=(189,249,255) rgb255=(235,255,255) rgb255=(255,255,235) rgb255=(255,242,189)
rgb255=(255,214,153) rgb255=(255,172,117) rgb255=(255,120,87) rgb255=(255,61,61)
rgb255=(247,40,54) rgb255=(217,22,48) rgb255=(166,0,33)}
}

\renewcommand{\emph}[1]{\textit{#1}}
\setenumerate{label=\textcolor{mybluegray1}{$\bm{\diamond}$},itemsep=0.0pt,topsep=5pt}

\pgfplotsset{every tick label/.append style={font=\scriptsize}}

\pgfplotsset{
	colormap/rdbur/.style={
		colormap={rdbur}{
			rgb255(0cm)=(20,46,97); 
			rgb255(1cm)=(52,100,171); 
			rgb255(2cm)=(83,146,194); 
			rgb255(3cm)=(153,197,222); 
			rgb255(4cm)=(211,229,240);
			rgb255(5cm)=(247,247,247);
			rgb255(6cm)=(250,219,200);
			rgb255(7cm)=(238,165,132);
			rgb255(8cm)=(207,94,80);
			rgb255(9cm)=(171,10,46);
			rgb255(10cm)=(99,0,32);
			}
		}
	}

\begin{document}

\preprint{Draft - 29032023-2106 } 

\title[Spatio-Temporal Learning of Turbulent Flows]{Autoregressive Transformers for Data-Driven Spatio-Temporal Learning of Turbulent Flows}

\author{Aakash Patil}

\author{Jonathan Viquerat} 
\email{jonathan.viquerat@minesparis.psl.eu}

\author{Elie Hachem}
\affiliation{%
MINES ParisTech, CEMEF\\
Paris Sciences and Lettres University\\
06904 Sophia Antipolis, France 
}%

\date{\today}

\begin{abstract}
A convolutional encoder-decoder-based transformer model is proposed for autoregressively training on spatio-temporal data of turbulent flows. The prediction of future fluid flow fields is based on the previously predicted fluid flow field to ensure long-term predictions without diverging. A combination of convolutional neural networks and transformer architecture is utilized to handle both the spatial and temporal dimensions of the data. To assess the performance of the model, a priori assessments are conducted, and significant agreements are found with the ground truth data. The a posteriori predictions, which are generated after a considerable number of simulation steps, exhibit predicted variances. The autoregressive training and prediction of a posteriori states are deemed crucial steps towards the development of more complex data-driven turbulence models and simulations. The highly nonlinear and chaotic dynamics of turbulent flows can be handled by the proposed model, and accurate predictions over long time horizons can be generated. Overall, the potential of using deep learning techniques to improve the accuracy and efficiency of turbulence modeling and simulation is demonstrated by this approach. The proposed model can be further optimized and extended to incorporate additional physics and boundary conditions, paving the way for more realistic simulations of complex fluid dynamics.
\end{abstract}

\keywords{Suggested keywords}
\maketitle

%

\def\figdir{./figures}

\section{Introduction}

The main factor in turbulent flows is convection, which makes tasks such as flow control and model reduction complex and challenging. These tasks become non-linear, high-dimensional, multi-scale, and non-convex optimization problems due to the dominance of convection over diffusion. Due to the vast amount of numerical and experimental data available for turbulent flows, data-driven approaches are now gaining popularity in the fluid mechanics community. These approaches use deep learning models to make predictions and represent a valid alternative to traditional methods. This article explores a new data-driven approach based on deep learning to estimate future fluid flow fields from previous ones. The proposed method uses a novel convolutional encoder-decoder transformer model and autoregressive training to achieve long-term spatio-temporal predictions. The approach is tested on two turbulent fluid flow cases, namely a wake-flow past a stationary obstacle and an environmental flow past a tower fixed on a surface. The results show the effectiveness of the proposed method in predicting the fluid flow fields accurately, highlighting the potential of data-driven approaches in solving challenging problems in fluid mechanics.  

There are several traditional ways to address temporal estimations, such as Koopman theory and proper orthogonal decomposition, which are suitable for prediction and control \cite{williams2015data,brunton2016koopman,rowley2017model}. Additionally, data assimilation schemes are popular, where the model weights are updated to reflect new observations \cite{mons2016reconstruction}. In recent years, supervised learning techniques using neural networks have been applied to capture nonlinear relations between past and future states. For example, a recurrent neural network with long-short term memory was used to predict the chaotic Lorenz system, and convolutional networks were used to predict transient flows \cite{dubois2020data,xu2020multi}. There have also been attempts to approximate the full Navier-Stokes equations using deep neural networks, but prediction accuracy decreased significantly for chaotic and turbulent flows \cite{lusch2018deep,sirignano2018dgm,tang2021exploratory,sun2020neupde}. Regarding the estimation of flow fields using deep neural networks, several studies have focused on spatial and temporal reconstruction, as well as spatial supersampling \cite{cheng2021deep,yousif2021high,schmidt2021machine}. Hybrid deep neural network architectures have been designed to capture the spatial-temporal features of unsteady flows \cite{han2019novel}, and machine learning-based reduced-order models have been proposed for three-dimensional complex flows \cite{nakamura2021convolutional}. A deep learning framework combining long short-term memory networks and convolutional neural networks has been used to predict the temporal evolution of turbulent flames \cite{ren2021predictive}. However, despite the significant progress made in the acceleration of flow simulation, these models still suffer from the generalization problem and are sensitive to parameter changes \cite{kochkov2021machine}. 

New deep learning architectures for temporal problems in unstructured and structured data are emerging, with transformers being one of the most promising. These models make use of self-attention mechanisms to differentially weight the significance of each part of the input data \cite{vaswani2017attention, bahdanau2014neural}, without the need for recurrent network architecture. Inspired by neighborhood-like notions in convolutional neural networks, transformers build features of inputs using a self-attention mechanism to determine the importance of other samples in the dataset with respect to the current sample. The updated features of the inputs are simply the sum of linear transformations of all features weighted by their importance. Transformers avoid recurrence by using the self-attention mechanism, which accounts for the similarity score between elements of a sequence and the positional embedding of these elements, allowing them to account for the full sequence instead of single elements. These models have been successful in natural language processing (NLP) tasks such as translation and text summarization, and are becoming the model of choice for NLP problems, replacing classical recurrent neural network (RNN) models such as long short-term memory (LSTM) \cite{wolf2020transformers, devlin2018bert,radford2019language}. Transformers have also been applied to image processing tasks using convolutional neural networks to capture relationships between different portions of an image \cite{dosovitskiy2020image, parmar2018image, touvron2021training}. Hybrid architectures combining convolutional layers with transformers have achieved excellent results in several computer vision tasks \cite{dai2021coatnet, wu2021cvt}. In spatio-temporal context, transformers have been used for video-understanding tasks, capturing spatial and temporal information through the use of divided space-time attention \cite{sharir2021image, bertasius2021space}. In fluid mechanics, attention mechanisms have enhanced the reduced-order model to extract temporal feature relationships from high-fidelity numerical solutions \cite{wu2021reduced}. Recently, a similar combination of autoregressive transformers and two-dimensional homogeneous isotropic turbulence was proposed for spatio-temporal prediction of flow fields \cite{peng2022attention}. However, transformers have never been used for spatio-temporal prediction of flow fields involving turbulent flows.

The present contribution is organized as follows: first, the deep learning method based on the convolutional self-attention transformer is discussed, after which focus is made on the autoregressive training procedure. The following section provides insights into the performance of the proposed approach by considering (i) a turbulent flow case with an obstacle embedded in a rectangular domain, and (ii) a surface-mounted tower in an open flow. This part is followed by a discussion and a conclusion.

\section{Deep Learning Method}
 
The primary focus of this contribution is to address the challenge of learning the spatio-temporal dynamics of turbulent flows, which are known for their high complexity, non-linear behavior, and high dimensionality. There are two main approaches to estimate a reference spatio-temporal field $X_t$: (i) reconstruction, which involves utilizing limited measurements $\tilde{X_t}$ at a specific time $t$ to reconstruct the full $X_t$ field at the same time, and (ii) prediction, where a dynamical model is utilized to advance the field in time based on previous estimates. Here, spatio-temporal learning is formulated as a task with a given time-series containing $N$ sequential snapshots $\left[ x_{t},x_{t+\Delta t}, ...., x_{t+(N-1)\Delta t} \right]$, in order to predict the same quantity of interest on $M$ steps ahead in time. The input $X$ of the deep learning model is $\left[ x_{t},x_{t+\Delta t}, ...., x_{t+(N-1)\Delta t} \right] $, and the output $Y$ is $\left[ x_{t+N\Delta t}, ..., x_{t+N+(M-1)\Delta t} \right]$. Each snapshot $x_{t}$ can be a scalar field or a vector field containing multiple features.  

Transformer models were created to address problems in natural language processing, such as completing sentences and translation by embedding words one by one. These tasks involve a sequence of words or sentence tensors measured over time and can be considered temporal learning problems \cite{vaswani2017attention}. Transformer models have demonstrated impressive results in a range of other tasks, including learning image patches as sequences, image reconstruction, and completion \cite{vaswani2017attention,dai2021coatnet, wu2021cvt}. As a result, transformer models are now challenging the traditional long short-term memory (LSTM) models, which are the \textit{de facto} RNNs, and are becoming the preferred state-of-the-art approach for a variety of temporal learning tasks.

Like RNNs, transformers are designed to handle sequential input data. However, unlike the latter, they do not necessarily process the data in order. Rather, the attention mechanism provides context for any position in the input sequence, and self-attention itself identifies/learns the weights of attention. In the case of spatio-temporal data, the attention can be applied to the spatial as well as the temporal sequence to attend to or pay attention to. The vanilla transformers in their original form are pure sequence to sequence models, as they learn a target output sequence from an input sequence, \textit{i.e.} they perform transformation at the sequence level. Their limitations, such as disrupting temporal coherence and failing to capture long-term dependencies, were reached for sentence completion of language generation tasks, where difficulties were noted while generating texts with a model which learns sequences without the knowledge of full-sequences \cite{bahdanau2014neural,yu2017seqgan, guo2018long}. Several studies were performed, such as that of Dai \textit{et al.} \cite{dai2019transformer}, to address this inability to capture long-term dependencies by attending to memories from previously learned parameters, yet at the expense of computing costs. To deal with some of these issues, autoregressive transformers were proposed by \cite{katharopoulos2020transformers} for sentence and image completion tasks. Although not explicitly stated in some works, the Generative Pre-trained Transformer (GPT) family of models \cite{radford2018improving, radford2019language, brown2020language} are in fact autoregressive transformers inspired by the decoder part of the original transformers. In \cite{katharopoulos2020transformers}, Katharopoulous \textit{et al.} showed that a self-attention layer trained in an autoregressive fashion can be seen as a recurrent neural network. Transformers can be combined with the classic convolutional encoder-decoder type models to harness their full potential when the input and target output tensors are in a spatio-temporal form. As locality is more important in learning small-scale features, this combination serves as a powerful method for a variety of computer-vision problems, including video-frame prediction. The self-attention mechanism on convolutional layers not only \textit{attends} or focuses on a sequence of significance, but it also improves the representation of spatially-relevant regions by focusing on important features and suppressing less-important ones \cite{woo2018cbam}.  

\begin{figure}
\centering
\includegraphics{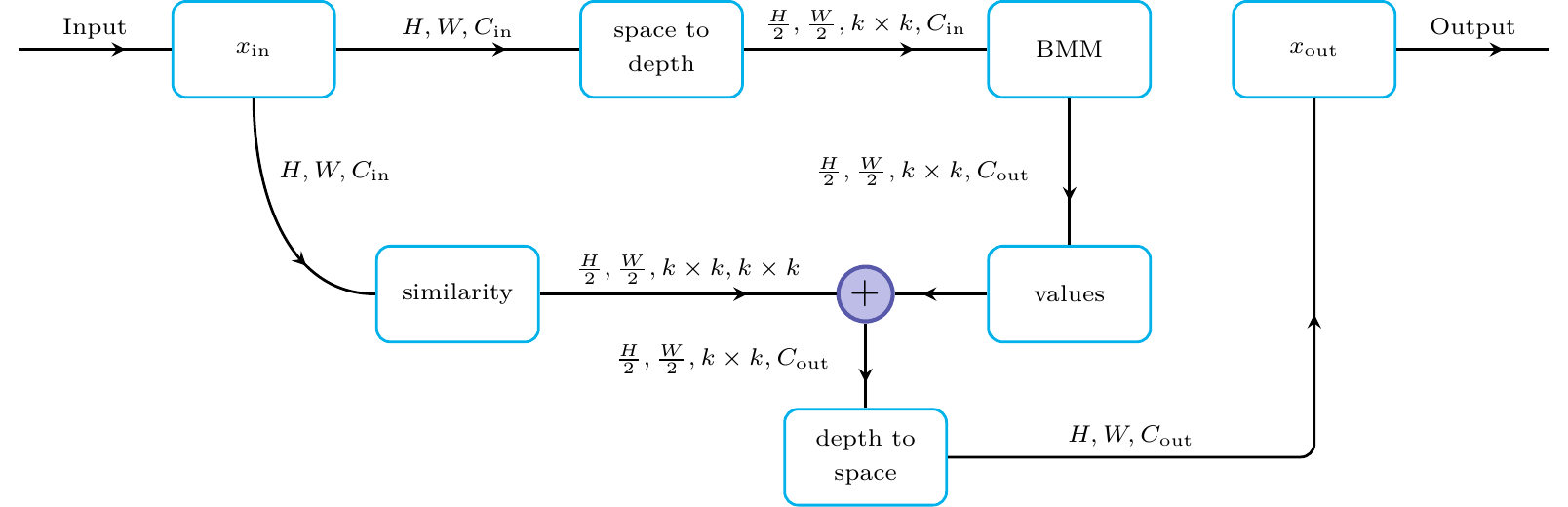}

\caption{\textbf{The convolutional transformer layer} is composed of two blocks: the batched matrix multiplication (BMM) and the self-attention summation. The BMM block corresponds to $W_{i \rightarrow j} x_{j}$ in equation (\ref{eqn:selfattend_layer}), with the batch dimension being the number of spatial locations. It performs $k \times k$ different input-dependent summations with the weights $\alpha$ in equation (\ref{eqn:selfattend_layer}). It contains both the learnable filter and the dynamic kernel.}
\label{fig:network_Conv}
\end{figure}

When a transformer block follows a convolutional layer, the model learns to highlight significant features across the channel sequence and spatial dimensions. The input sequences are initially concatenated channel-wise to the input layer, and subsequent convolutional operations take place in the encoder. In the convolutional layers, the intermediate feature maps \(\mathbb{F}\in \mathbb{R}^{C\times H\times W}\) from a specific layer go through the self-attention convolutional transformer layer $\text{conv}_{\alpha}$, which attends to both spatial representation and the positional embeddings of the input sequence channels. In $\text{conv}_{\alpha}$, let $x, y \in \mathbb{R}^{C}$ denote the input and output intermediate feature tensors, where $C$ indicates the number of intermediate channels. When $i,j \in \mathbb{R}^{H\times W}$ are indices of the spatial nodes, a standard convolution operation occurs as follows:
\begin{equation}
    \begin{split}
    y_{i} = \sum_{j \in N(i)}{W_{i \rightarrow j} x_{j}},\\
    \end{split}
    \label{eqn:conv_layer}
\end{equation}
\noindent where $N(i)$ signifies the spatial nodes in a local neighborhood defined by a kernel of size $k \times k$ centered at node $i$, $i \rightarrow j$ denotes the relative spatial relationship from $i$ to $j$, and $W_{i \rightarrow j} \in \mathbb{R}^{C \times C}$ is the weight matrix.
On the other hand, self-attention for intermediate convolutional features has three weight matrices $W_{q}, W_{k}, W_{v} \in \mathbb{R}^{C \times C}$ to compute query, key and value respectively. For each convolution window, the self-attention is given as:

\begin{equation}
    \begin{split}
    y_{i} &= \sum_{j \in N(i)}{\alpha_{i \rightarrow j} W_{v} x_{j}},\\
    \alpha_{i \rightarrow j} &= \frac{e^{(W_{q}x_{i})^{T}W_{k}x_{j}}}{\sum_{z \in N(i)}{e^{(W_{q}x_{i})^{T}W_{k}x_{z}}}} = \frac{W_{qk}x_{i}[j]}{\sum_{z \in N(i)}{W_{qk}x_{i}[z]}},
    \end{split}
    \label{eqn:selfattend_layer}
\end{equation}

\noindent where the self-attention $\alpha_{i \rightarrow j} \in (0, 1)$ is a scalar that controls the contribution of values in spatial nodes, with $W_{qk} \in \mathbb{R}^{C \times k^{2}}$, and $[j]$ means $j^{th}$ element of the tensor. $\alpha$ is usually normalized by a softmax operation such that $\sum_{j}{\alpha_{i \rightarrow j}} = 1$. These operations are summarized in figure \ref{fig:network_Conv}.  

Combining equations (\ref{eqn:conv_layer}) and (\ref{eqn:selfattend_layer}), one obtains both an input sequence dependent kernel and the learnable convolution filters providing the final output feature map  \(\mathbb{F''}\) by convolutional transformer layer, given as:

\begin{equation}
    \begin{split}
    y_{i} &=  \sum_{j \in N(i)}{ \operatorname{softmax}\left(\alpha_{i \rightarrow j}\right) W_{i \rightarrow j}} x_{j}  \\
    \textit{i.e.} \,\, \mathbb{F''} &= \text{conv}_{\alpha}(\mathbb{F})  
    \end{split}
\end{equation}
 
The current self-attention convolutional transformer layer has a $3 \times 3$ kernel and incorporates the representation of convolutional features. Combining convolutional neural networks with self-attention thus offers superior learning capabilities of spatio-temporal structures, which would benefit turbulent flows and CFD in general, where one learns spatial filters as well as temporal embeddings and dependencies. In addition to the convolutional transformer layer, the model is trained in an autoregressive fashion. Formally, autoregressive models are those which forecast future sequences from the previously forecasted sequences in a cyclical way, and thus here \emph{auto} indicates the regression of the variable sequence against itself.

In turbulent flow problems, the high-dimensional state-space is characterized by intricate spatio-temporal dynamics, and therefore, dimensionality reduction techniques can be useful \cite{dubois2022machine}. The prediction and reconstruction problems can be interpreted as estimating the reduced or latent state, making it natural to use encoder-decoder architectures. The encoder takes input tensors, learns the most relevant parts, and maps them to a high-dimensional representation. This high-dimensional representation is then converted to target output tensors by the decoder, which involves successive up-samplings and convolutions. By connecting the encoder and decoder, their weight matrices learn to jointly map the input to the output tensors, allowing small-scale features to be learned. The decoder aims to transform the latent space representation with a dimension of $n_z \times n_z$ to the original spatial dimensions of the target output at time $x_{t+\Delta t}$.

\begin{figure}
\centering
\includegraphics{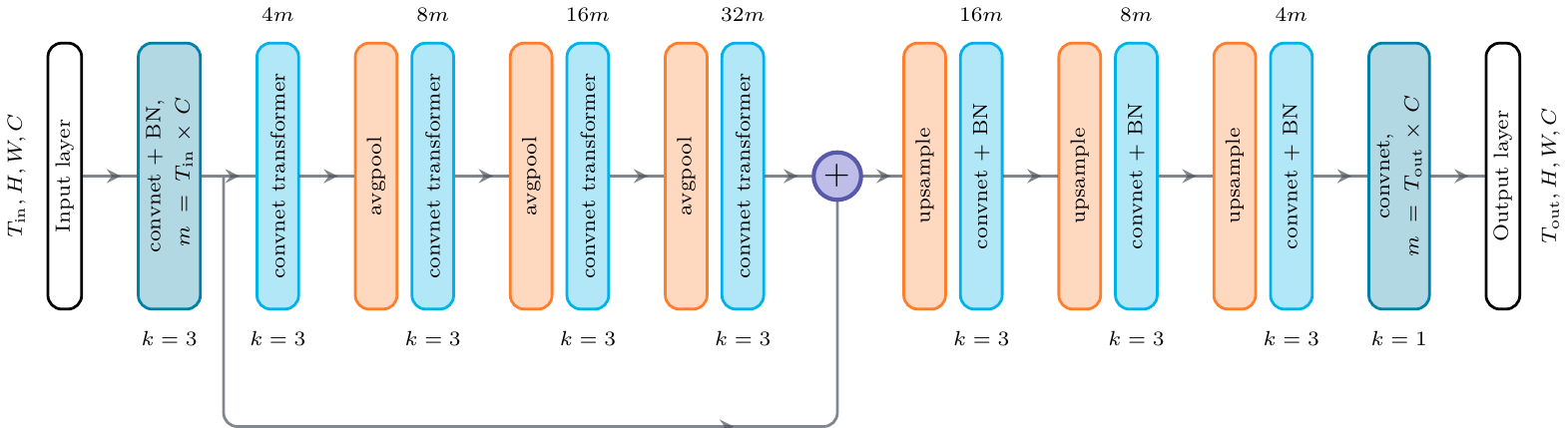}

\caption{\textbf{Convolutional encoder-decoder transformer architecture } Model architecture of the convolutional encoder-decoder transformer to process low and high level features. The canonical four-stage design is utilized in addition to the convolutional transformer blocks or layers. $H,W$ are the input resolutions for each snapshot in $T_{\text{in}}$ sequence and $T_{\text{out}}$ sequence.}
\label{fig:network}
\end{figure}

For a trained model $\mathbb{M}$ as shown in figure \ref{fig:network}, multi-step training is performed for quantity $X_{t}$ in an auto-regressive manner, \textit{i.e.} $X_{t+\Delta t}$ is predicted from previously predicted $X_{t}$, where $t$ is some non-dimensional time. In other words, an initial condition $X_{t}$ is inputted to the model to learn $\widehat{X}_{t+\Delta t}$, after what this predicted $\widehat{X}_{t+\Delta t}$ is then fed back to the model again to learn $\widehat{X}_{t+2\Delta t}$ and so on, in an autoregressive manner:  

\begin{equation}
	\left\{
	\begin{split}
		\widehat{X}_{t+\Delta t} 	&= \mathbb{M}(X_{t}), \\
		\widehat{X}_{t+2\Delta t} 	&= \mathbb{M}(\widehat{X}_{t+\Delta t}), \\
							&\,\, ... \\
		\widehat{X}_{t+(n-1)\Delta t} 	&= \mathbb{M}(\widehat{X}_{t+(n-2)\Delta t}), \\
	\end{split}
	\right.
\end{equation}

\noindent where $t$ is the time step and $X \in \mathbb{R}^{C \times H \times W}$ the input tensor snapshot at instant $t$. In the following, the autoregressive training sequence length is set equal to two in order to limit the computational cost. 

In order to preserve meaningful values at the boundaries, the convolutional filters used in the proposed architecture incorporate a symmetric boundary condition into the padding operation. While padding is typically used to retain the spatial dimensions of the field undergoing convolution, zero-padding does not accurately represent the expected physical behavior.  Indeed, padding with zeros everywhere would violate the representation of existing boundary conditions, for example, the notion of wall boundaries would have lesser significance if a region is padded with zeros on all the sides in a channel flow. To preserve the boundary conditions after multiple successive convolutions, a boundary condition formulation was implemented such that the walls could be padded with zeros if required, while the other sides could be padded with adequate values from the symmetric cells. To train the model, the Adam\cite{kingma2014adam} optimizer is used to iteratively minimize the total equi-weighted mean squared error (MSE) loss defined by:

\begin{equation}
	\begin{split}
		\mathcal{L} = \frac{1}{n_s} \biggr[ \sum_{i=1}^{n_s} \left( \left(X_{t+\Delta t} \right)^{i}  - \left(\widehat{X}_{t+\Delta t} \right)^{i} \right)^2 	
								&+ \sum_{i=1}^{n_s} \left( \left(X_{t+2\Delta t} \right)^{i}  - \left(\widehat{X}_{t+2\Delta t}\right)^{i} \right)^2 + \, \cdots \\
								& + \sum_{i=1}^{n_s} \left( \left(X_{t+(n-1)\Delta t} \right)^{i}  - \left(\widehat{X}_{t+(n-1)\Delta t}\right)^{i} \right)^2 \biggr].
	\end{split}
\label{lossRelation}
\end{equation}
 
The activation function used in the neural network was ReLU, which is known to help stabilize the weight update during training \cite{nair2010rectified}. During training, the entire training dataset was presented to the network repeatedly after shuffling, and each complete pass is called an epoch. An early stopping criterion was used to stop the training process, along with a learning rate reduction if learning improvement did not occur after every 100 epochs. The TensorFlow platform \cite{abadi2016tensorflow} was used to implement the neural network, and Nvidia Tesla V100 GPUs were used to train it.

\section{Numerical simulation cases and data generation} 

\subsection{Governing equations}
\label{section:equations}

The evolution of the velocity $\boldsymbol{u}$ and pressure $p$ in an incompressible fluid flow is governed by the Navier-Stokes momentum and continuity equations, given with positive constant density $\rho$ and dynamic viscosity $\mu$ as :

\begin{equation} \label{eq:ns_equation1}
	\begin{split}
		\rho \left( \partial_{t} \boldsymbol{u} + \boldsymbol{u} \cdot \mathbf{\nabla} \boldsymbol{u} \right) &= \mathbf{\nabla} \cdot \left( - p\;{\mathbf{I}_d} + 2 \mu\; \boldsymbol{\varepsilon}( \boldsymbol{u} ) \right)  + \boldsymbol{f}, \\
													\mathbf{\nabla} \cdot \boldsymbol{u} &= 0,
	\end{split}
\end{equation}

\noindent Equation (\ref{eq:ns_equation1}) includes the strain-rate tensor, $\boldsymbol{\varepsilon}(\boldsymbol{u})$, the identity tensor ${\mathbf{I}d}$, and an additional forcing or source term $\boldsymbol{f}$. To solve a physical problem, suitable boundary and initial conditions are added to the equations. The presence of turbulence is accounted for by including an eddy viscosity term $\mu{t}$ in the equations, which is modeled based on one or more turbulent scales. The eddy viscosity is computed using the Spalart-Allmaras (SA) model, which is a one-equation model that solves a convection-diffusion-reaction problem to describe the evolution of the kinematic eddy viscosity-like variable $\nu_{t}$ \cite{spalart1992one}.

\subsection{Case 1 : Wake-flow past a square cylinder}

The turbulent flow around a 2D square cylinder is examined as a widely used benchmark case. The reference velocity $U_{\infty}$ and cylinder lateral size $H$ at the domain's center are used to set the baseline Reynolds number to \num{22e3}. The computational domain spans $\left[ -5H, 15H \right] \times \left[ -7H, 7H \right]$ in the streamwise $x$ and crosswise $y$ directions, respectively. To conduct a URANS or VMS simulation, the domain is discretized into a sufficient number of cells using a finite-element flow solver developed in-house \cite{bazilevs2007variational, takizawa2018stabilization, hachem2013immersed, guiza2020anisotropic}.  The inflow boundary conditions comprise of ${ u}=(V_{in},0)$ and $\tilde{\nu}=3\nu$, leading to an eddy to kinematic viscosity ratio of about $0.2$. The side boundaries are treated as symmetrical, with $\partial_y u_{x}=u_{y}=0 $ and $\partial_y \tilde{\nu}=0 $. For the outflow, $\partial_x u_{x}= \partial_x u_{y}=0,$ $\partial_x \tilde{\nu}=0$ along with $p=0$ are enforced. At the cylinder surface, no-slip conditions ${ u}=0 $ and $ \tilde{\nu}=0$ are applied. The simulation runs for a physical time of $5000$ seconds with a time step of $\Delta t = 0.05 $ seconds. A stable flow is achieved after about 200 seconds, and the remaining data for the next 4800 seconds is collected for training and testing purposes. Data is recorded every $\Delta t = 0.25 $ seconds, resulting in about 1500 snapshots. In terms of non-dimensional time defined as $t^{*} = t U_{\infty}/H$, this sampling rate corresponds to $\Delta t^{*} = 1 $. Approximately 24 shedding cycles are observed in simulation data. Given the 70/30 splitting strategy, 16 shedding cycles are observed in training data, which seems reasonable to fully characterize the dynamics of wake turbulent flow past a two-dimensional square cylinder considering its simplicity. Figure \label{fig:cylinder_sketch}  shows the sketch of the associated case. 
 
\begin{figure}
\centering
\begin{subfigure}[t]{.8\textwidth}
	\centering
    \includegraphics{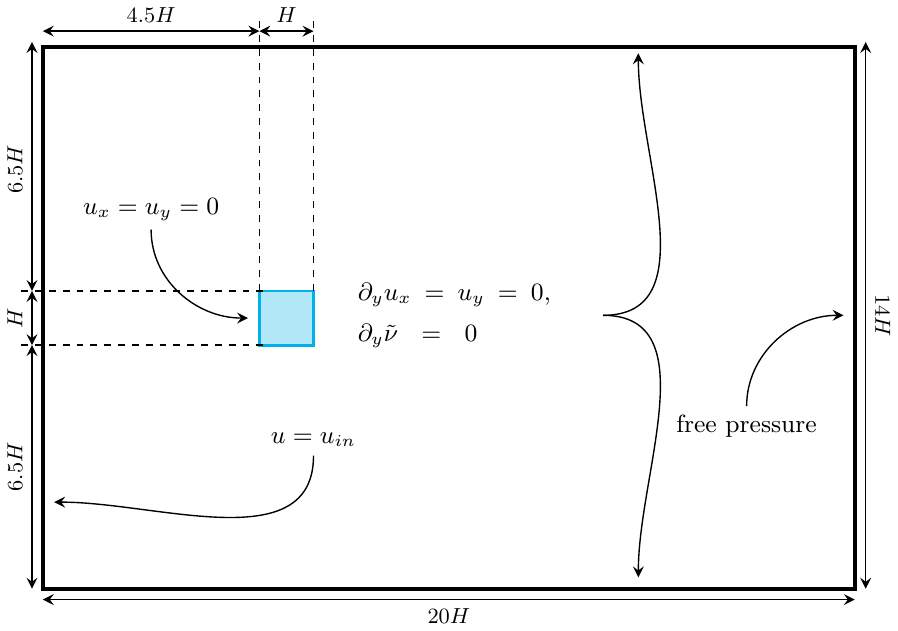}
	\caption{ }
	\label{fig:cylinder_sketch}
\end{subfigure}
\medskip
\begin{subfigure}[t]{.8\textwidth}
	\centering
    \includegraphics{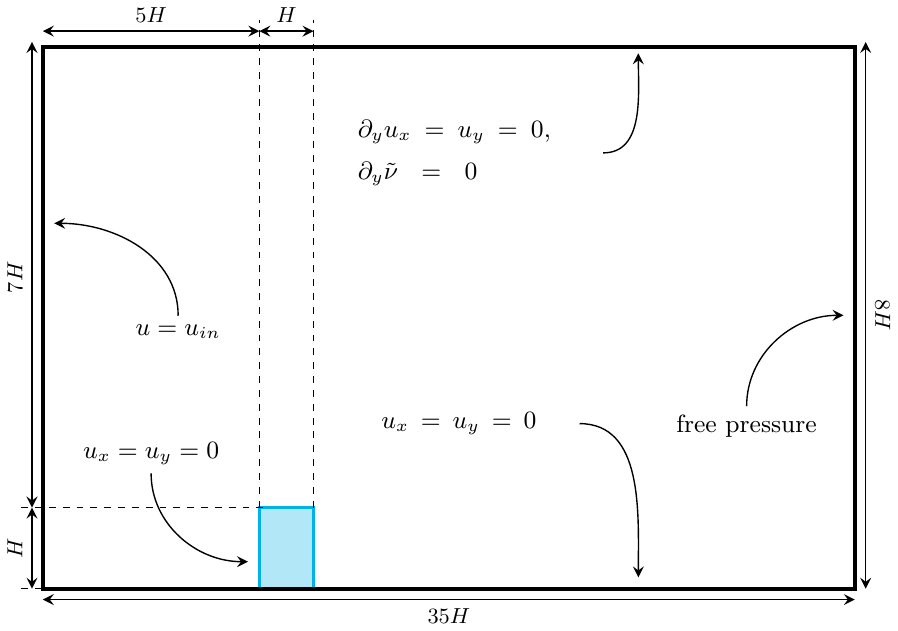}
	\caption{ }
	\label{fig:tower_sketch}
\end{subfigure}
\caption{(a) Case 1 setup: Wake-flow past a square cylinder (b) Case 2 setup: Environmental flow over a surface-mounted tower}
\end{figure}
 
\subsection{Case 2 : Environmental flow over surface-mounted tower}

The turbulent flow past a two-dimensional (2D) rectangular tower on the land surface is considered.  The baseline Reynolds number is set to \num{45e2}, based on the reference velocity $U_{\infty}$ and the square tower of sides $H$ which is placed on the surface. The dimensions of the computational domain are $\left[ -5H, 30H \right] \times \left[-H, 7H \right]$ in the streamwise $x$ and crosswise $y$ directions respectively, and the domain is discretized into sufficiently large number of cells to perform a URANS or VMS simulation. The inflow boundary conditions are ${  u}=(V_{in},0)$, together with $\tilde{\nu}=3\nu$, which corresponds to a ratio of eddy to kinematic viscosity of approximately $0.2$. For the top of the domain, the velocity component normal to the surface is set to zero.  No-slip boundary conditions ${ u}=0 $ and $ \tilde{\nu}=0$ are imposed at the tower surface, as well as the bottom surface at $y= -1$. The time step is $\Delta t = 0.01 $ seconds and $300$ seconds are simulated. For a statistically steady state to be reached (periodic vortex shedding to be observed), around $100$ seconds are required. The data of the remaining 200 seconds (\textit{i.e.} approximately \num{20e3} time steps) is stored for training and testing purposes. The data is sampled at each $\Delta t = 0.1 $ seconds, thus collecting around 2000 snapshots. In terms of non-dimensional time defined as $t^{*} = t U_{\infty}/H$, this sampling at each $\Delta t = 0.1 $ denotes $\Delta t^{*} = 1 $. Figure \ref{fig:tower_sketch} shows the sketch of the associated case.  Initially, a free separated shear layer expands above the tower and becomes wavy, and then reattaches at the bottom surface of the domain. The shear layer flaps and vortical structures are shed from it. Approximately 18 shedding cycles are observed in simulation data. Given the 70/30 splitting strategy, 12 shedding cycles are observed in training data enough to reasonably characterize the dynamics of environmental flow over the surface-mounted obstacle. 

\section{Results and Discussion}

\begin{figure}
\centering

    \includegraphics{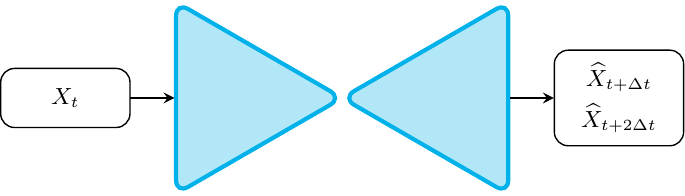}
    \includegraphics{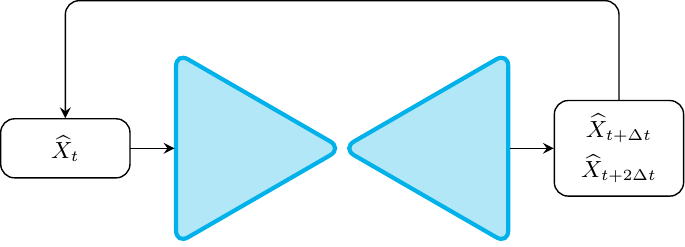}

\caption{\textbf{Illustration of \textit{a priori} and \textit{a posteriori} simulations} Left: For \textit{a priori} simulation, each $X_{t}$ from $\{X\}$, the dataset not used in training time, is fed to the model. Right: For \textit{a posterioi} simulation, the inputs $\widehat{X_{t}}$ are received from its own previous predictions, provided it was initiated with a suitable $X_{t}$.}
\label{fig:figure_apri_apost}
\end{figure}

\begin{figure}
\centering
\begin{subfigure}[b]{.45\linewidth}
	\centering
    \includegraphics{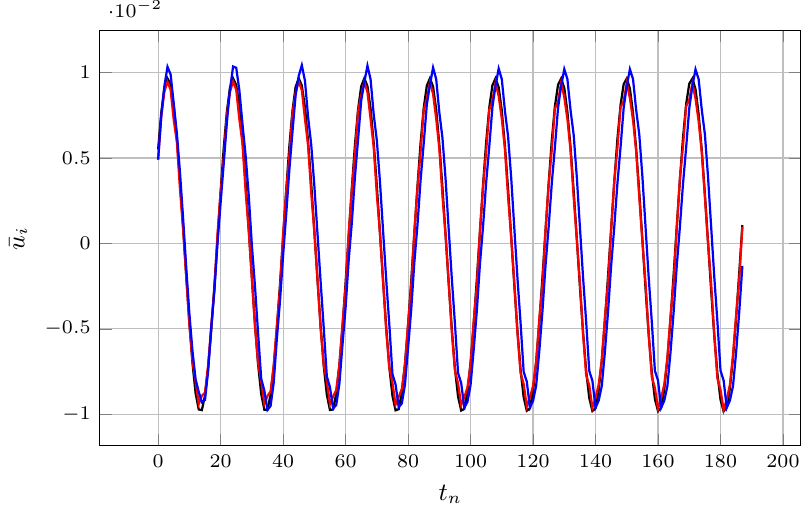}
	\caption{\textit{A priori} and \textit{a posteriori} predictions compared to ground truth for case 1}
	\label{fig:tDevCase1}
\end{subfigure} \quad
\begin{subfigure}[b]{.45\linewidth}
	\centering
    \includegraphics{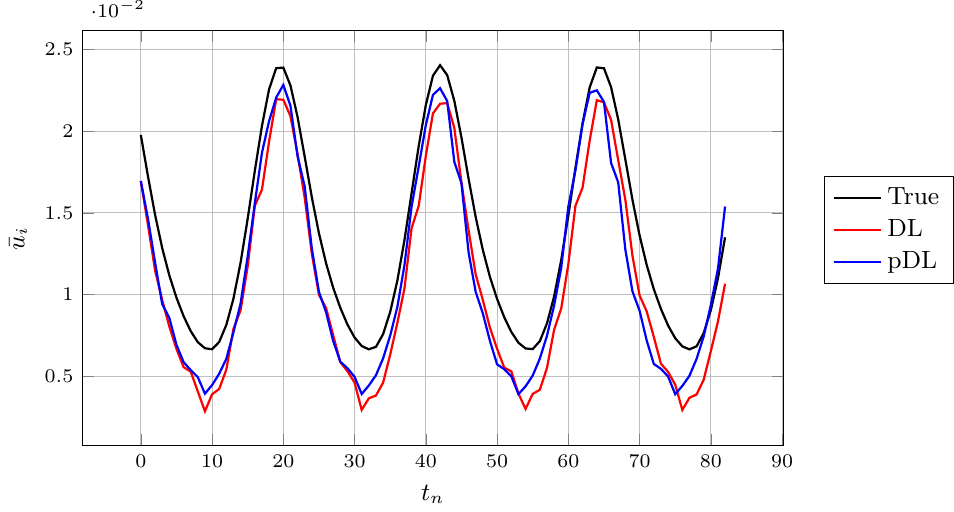}
	\caption{\textit{A priori} and \textit{a posteriori} predictions compared to ground truth for case 2}
	\label{fig:tDevCase2}
\end{subfigure}
\caption{\textbf{Temporal evolution of the ensemble averages} for \textit{a priori} and \textit{a posteriori} values of velocity magnitude compared to the true values in black. Left: Ensemble mean for spatial values of velocity magnitude for case 1. Right:  Ensemble mean for spatial values of velocity magnitude for case 2.} 
\label{fig:timeDeviations_velMag}
\end{figure}

In this section, the results are discussed as follows: first, the temporal evolutions of the quantities are compared, then the spatial measurements at various times are compared for velocity components. In a second time, temporal propagation of errors and correlation coefficients are compared along with the propagation of phase shifts. Additionally, the contour plots of quantities are also compared to provide qualitative assessments. These comparisons are performed for both the cases and for both the \textit{a priori} and \textit{a posteriori} simulations as illustrated in figure \ref{fig:figure_apri_apost}. To compare results, a first \textit{a priori} simulation is performed by exploiting data samples that were not used during training. The trained model is fed snapshots at instant $t$, and predicts the next two snapshots at instants $t+\Delta t$ and $t+2\Delta t$, and the process is repeated by feeding the subsequent snapshots from the dataset, until the same number of snapshots is reached for comparison with the original ground truth time series. As snapshots from the dataset are utilized, this approach is termed \textit{a priori} deep learning simulation. On the other hand, \textit{a posteriori} simulation is performed by feeding a snapshot at instant $t_0$ from the same dataset not used during training, and by predicting the next two snapshots at instants $t+\Delta t$ and $t+2\Delta t$. This predicted snapshot at instant $t+2\Delta t$ is then injected back into the model to predict snapshots at instants $t+3\Delta t$ and $t+4\Delta t$, and the process is similarly repeated until the same number of snapshots are obtained so as to compare with the true snapshots. This way of recycling the model predictions is termed \textit{a posteriori} deep learning simulation. Once an equal length of time snapshots are obtained, both the \textit{a priori} and the \textit{a posteriori} results against the truth from the dataset can now be compared. Figure \ref{fig:timeDeviations_velMag} shows the temporal evolution of the ensemble average of velocity magnitude for case 1 and case 2. For case 1, both \textit{a priori} and \textit{a posteriori} predictions present a good agreement with respect to the truth, whereas for case 2 the predictions, though fairly accurate, suffer from deterioration. Moreover, the long-term predictions of the model are evident from the accuracy of a posteriori predictions, giving us an indication of global long-term learning while comparing ensemble averages.

\begin{figure}
\centering
\begin{subfigure}[t]{.8\textwidth}
	\centering
    \includegraphics{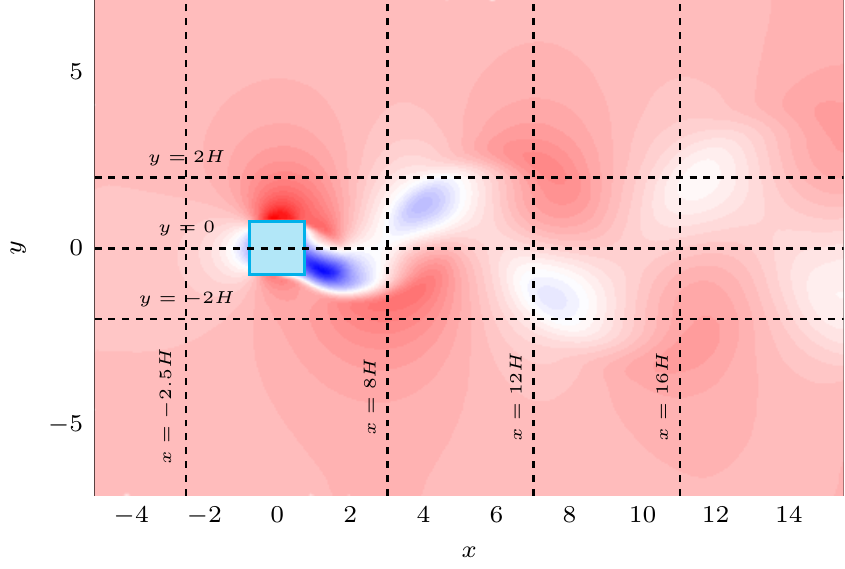}
	\caption{ }
	\label{fig:probeLocCase1}
\end{subfigure}

\medskip
\begin{subfigure}[t]{.8\textwidth}
	\centering
    \includegraphics{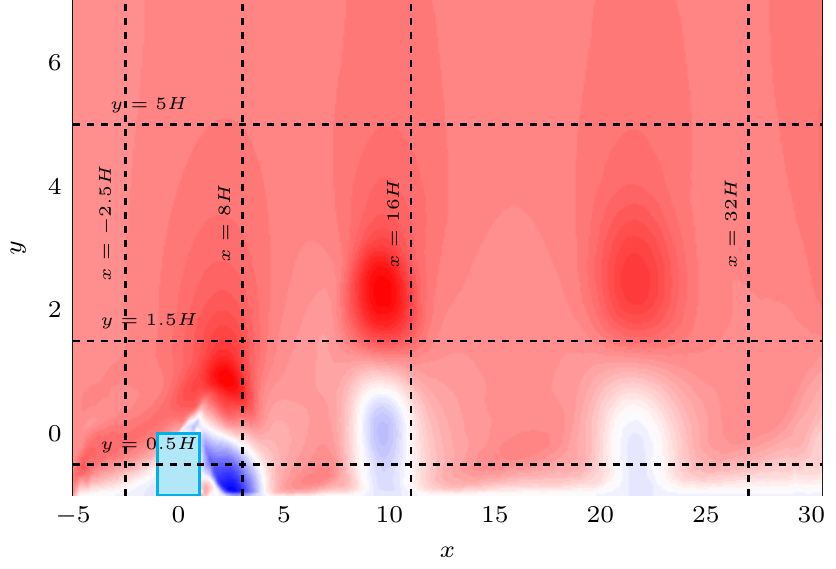}
	\caption{}
	\label{fig:probeLocCase2}
\end{subfigure}
\caption{\textbf{Locations of the probe lines used for comparison to the reference quantities.} Top: Lines along streamsie and cross-streamwise directions for case 1. Bottom: For case 2. }
\label{fig:probelines_alongXY}
\end{figure}

The accuracy of the predictions is further verified by comparing the values along the various streamwise and cross-streamwise locations. These locations are marked with dashed lines in figure \ref{fig:probelines_alongXY} for both the cases. For case 1, measurements were made along streamwise directions at $x=[ -2.5H, 8H, 12H, 16H ]$ and cross-streamwise directions at $y={ -2H, 0, 2H }$, and similarly for case 2, the measurements were made  at $x=[ -2.5H, 8H, 16H, 32H ]$ and at $y=[ 0.5H, 1.5H, 5H ]$. As the wake-flows are topic of interest, these locations were chosen based on the region of interest away from the obstacle for both the cases. With regards to temporal evolution, the predictions were compared at a certain percentage of the total predicted snapshots. As a reminder, around 200 snapshots were predicted for case 1 and around 100 snapshots for case 2. The predictions are compared at instants $t=[ 2\%, 33\%, 66\% ]$ to verify the quality of temporal evolution. 

\begin{figure} 
\centering

\begin{subfigure}[t]{.9\linewidth}
	\centering
     \includegraphics{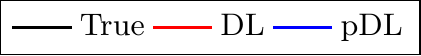}
	\caption*{}
\end{subfigure}
\vspace{-1em}%

\begin{subfigure}[b]{\linewidth}
\centering
\begin{subfigure}[b]{.4\linewidth}
     \includegraphics{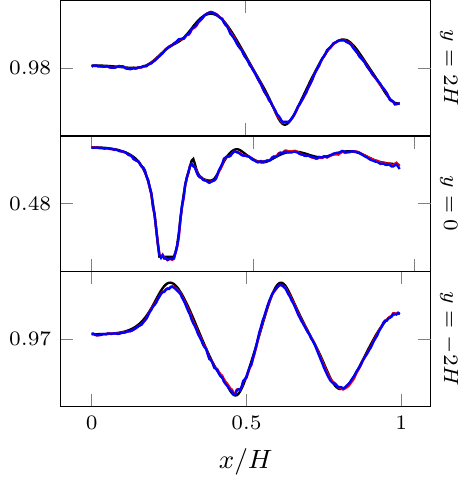}
\end{subfigure} \quad
\begin{subfigure}[b]{.4\linewidth}
      \includegraphics{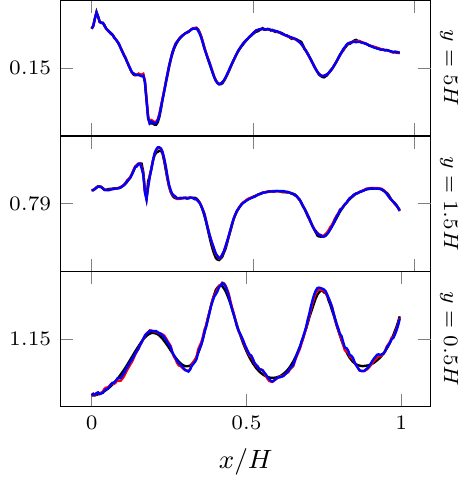}
\end{subfigure}  
\caption{Time $t= 0.02\, T_n$ }
\end{subfigure} 

\medskip
\medskip

\centering
\begin{subfigure}[b]{\linewidth}
\centering
\begin{subfigure}[b]{.4\linewidth}
     \includegraphics{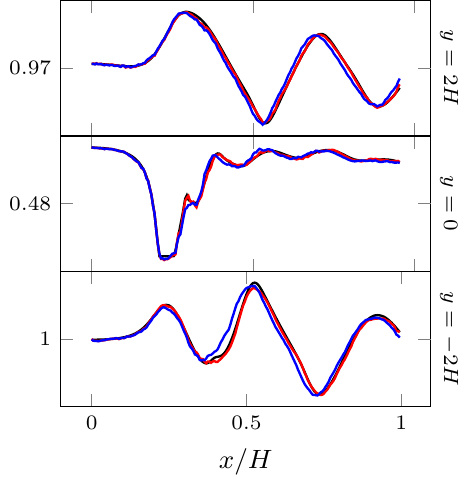}
\end{subfigure} \quad
\begin{subfigure}[b]{.4\linewidth}
      \includegraphics{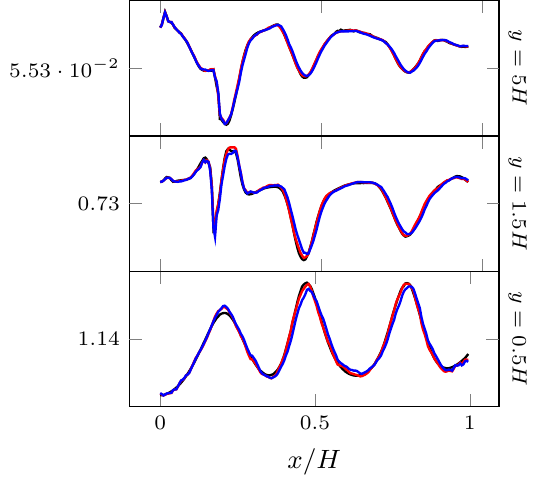}
\end{subfigure} 
\caption{Time $t= 0.33\, T_n$ }
\end{subfigure} 

\medskip
\medskip

\centering
\begin{subfigure}[b]{\linewidth}
\centering
\begin{subfigure}[b]{.4\linewidth}
      \includegraphics{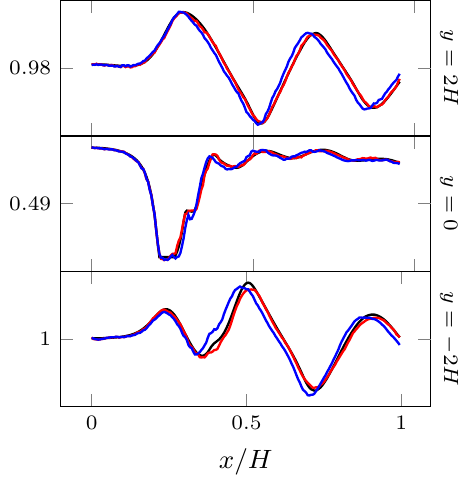}
\end{subfigure} \quad
\begin{subfigure}[b]{.4\linewidth}
         \includegraphics{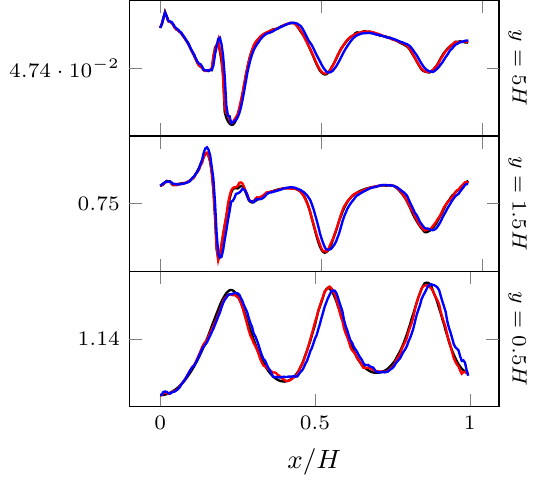}
\end{subfigure} 
\caption{Time $t= 0.66\, T_n$ }
\end{subfigure} 

\caption{\textbf{Comparative predictions of streamwise velocity components ($u_0$) sampled along y-axis for case 1 (left) and case 2 (right).} Figures from top to bottom denote the predictions at increasing times, \textit{i.e.} the top row contains instantaneous preditctions at $t= 0.02\, T_n$,  the middle row at $t= 0.33\, T_n$, and the bottom row shows the predictions at $t= 0.66\, T_n$ . }
\label{fig:comparison_pripost_TU_aY}
\end{figure}

\begin{figure} 
\centering

\begin{subfigure}[t]{.9\linewidth}
	\centering
     \includegraphics{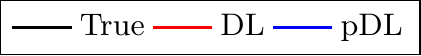}
	\caption*{}
\end{subfigure}
\vspace{-1em}%

\begin{subfigure}[b]{\linewidth}
\centering
\begin{subfigure}[b]{.4\linewidth}
     \includegraphics{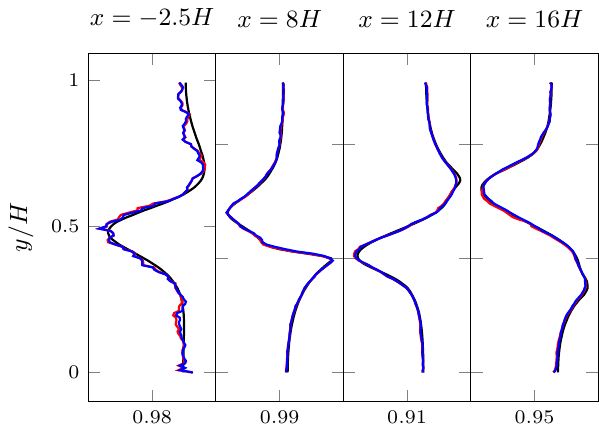}
\end{subfigure} \quad
\begin{subfigure}[b]{.4\linewidth}
      \includegraphics{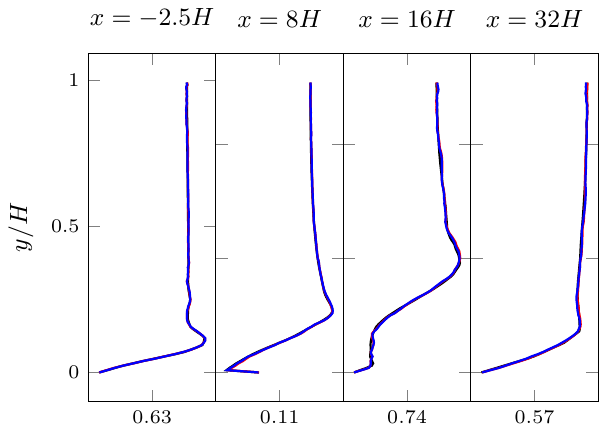}
\end{subfigure}  
\caption{Time $t= 0.02\, T_n$ }
\end{subfigure} 

\medskip
\medskip

\centering
\begin{subfigure}[b]{\linewidth}
\centering
\begin{subfigure}[b]{.4\linewidth}
     \includegraphics{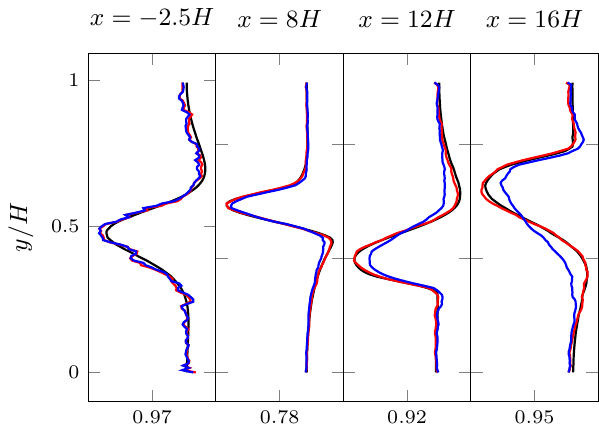}
\end{subfigure} \quad
\begin{subfigure}[b]{.4\linewidth}
      \includegraphics{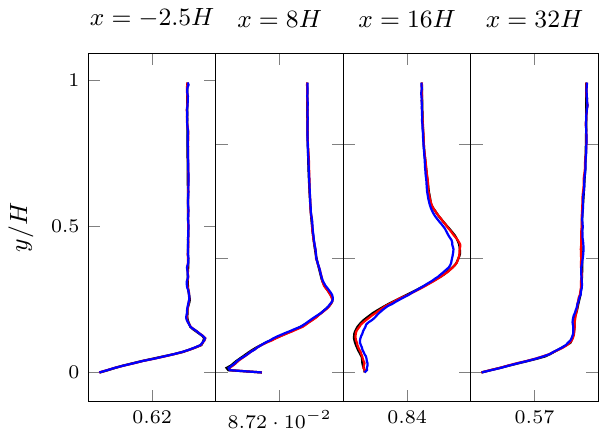}
\end{subfigure} 
\caption{Time $t= 0.33\, T_n$ }
\end{subfigure} 

\medskip
\medskip

\centering
\begin{subfigure}[b]{\linewidth}
\centering
\begin{subfigure}[b]{.4\linewidth}
      \includegraphics{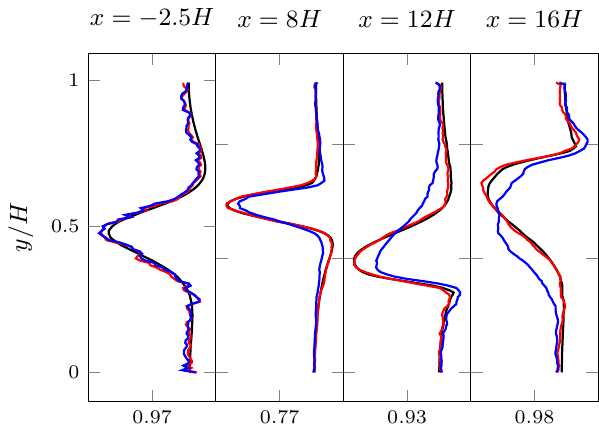}
\end{subfigure} \quad
\begin{subfigure}[b]{.4\linewidth}
         \includegraphics{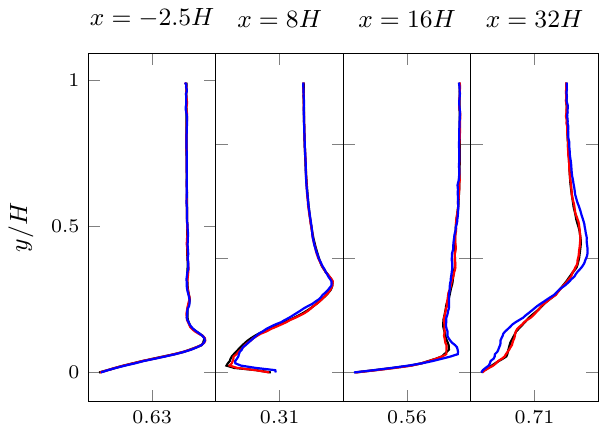}
\end{subfigure} 
\caption{Time $t= 0.66\, T_n$ }
\end{subfigure} 

\caption{\textbf{Comparative predictions of streamwise velocity components ($u_0$) for case 1 (left) and case 2 (right).} Figures from top to bottom, denote the predictions at increasing times, \textit{i.e.} the top row contains instantaneous predictions at $t= 0.02\, T_n$,  the middle row at $t= 0.33\, T_n$, and the bottom row shows the predictions at $t= 0.66\, T_n$.}
\label{fig:comparison_pripost_TU}
\end{figure}

Figure \ref{fig:comparison_pripost_TU} shows the evolution of temporal predictions of streamwise velocity component $u_0$ when measured along with cross-streamwise directions. The \textit{a priori} predictions follow closely the reference values indicating a good agreement with the short-term predictions along with the measured spatial directions. For \textit{a posteriori} predictions, an increasing deviation from the reference was observed as time evolves, which can be attributed to the accumulation error while making long-term predictions. Similarly, the evolution of the same quantity ($u_0$) when measured along streamwise directions is shown in figure \ref{fig:comparison_pripost_TU_aY}. A similar trend is observed for the predictions against the reference, where the a posteriori predictions deteriorate as time evolves. It can be noted that the upstream predictions at $x=-2.5H$ are better across times, as it is not affected by the turbulent wake. Overall measurements indicate a decent agreement of both the short-term \textit{a priori} predictions as well as long-term \textit{a posteriori} predictions with the reference solutions.

\begin{figure}
\centering
\begin{subfigure}[b]{.45\linewidth}
	\centering
         \includegraphics{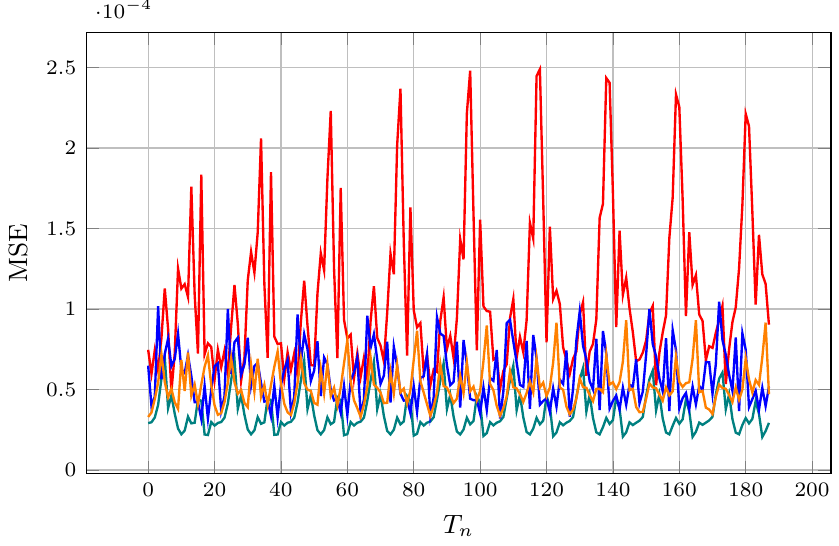}
	\caption{ }
\end{subfigure} \quad
\begin{subfigure}[b]{.45\linewidth}
	\centering
         \includegraphics{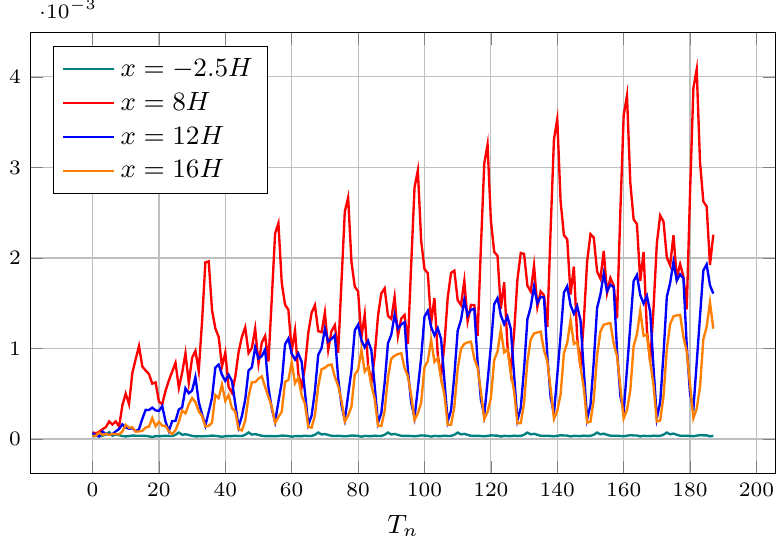}
	\caption{}
\end{subfigure}

\begin{subfigure}[b]{.45\linewidth}
	\centering
         \includegraphics{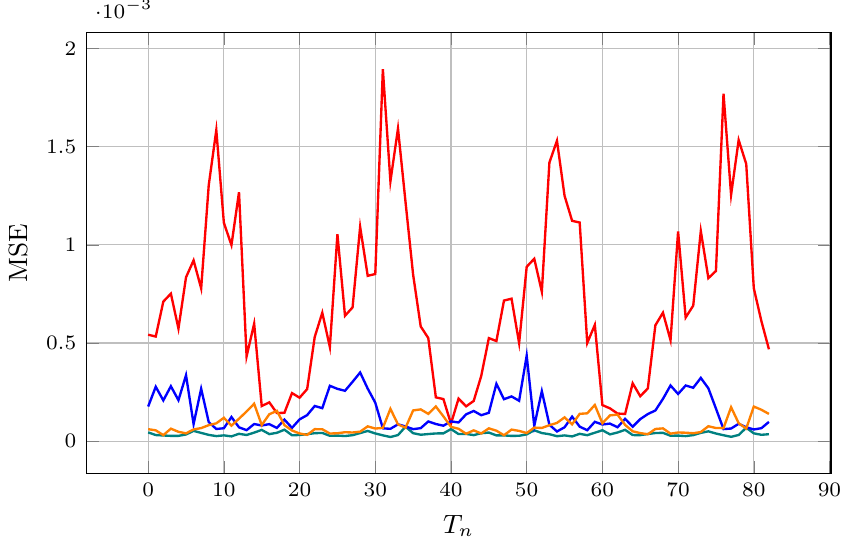}
	\caption{ }
\end{subfigure} \quad
\begin{subfigure}[b]{.45\linewidth}
	\centering
         \includegraphics{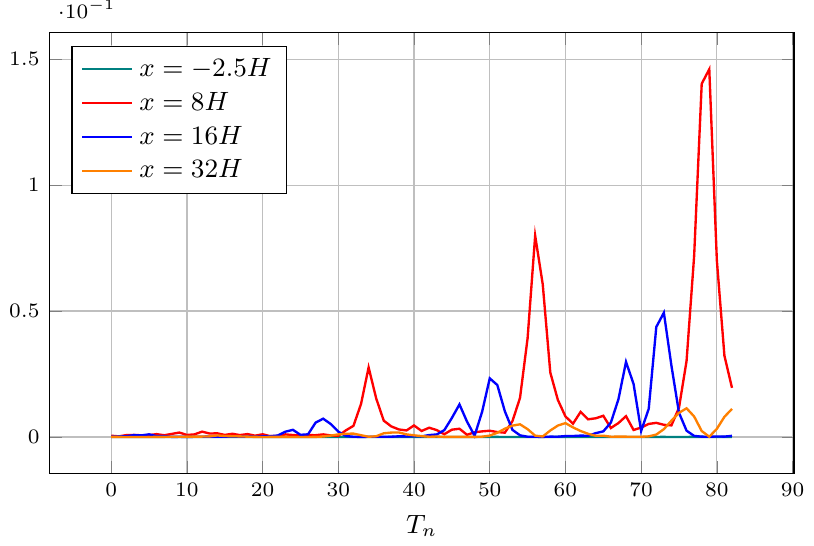}
	\caption{}
	\label{fig:tDevCase2}
\end{subfigure}

\caption{\textbf{Mean squared error propagation} for velocity magnitude with respect to the reference values. The top row shows the evolution for case 1, and the bottom row shows the evolution for case 2. On the left is the temporal evolution of a priori mean squared error. While on the right are the temporal evolution of a posteriori mean squared error. The values are shown for locations along the X-axis.} 
\label{fig:timeDistance_velMag}

\end{figure}


\begin{figure}
\centering
\begin{subfigure}[b]{.45\linewidth}
	\centering
         \includegraphics{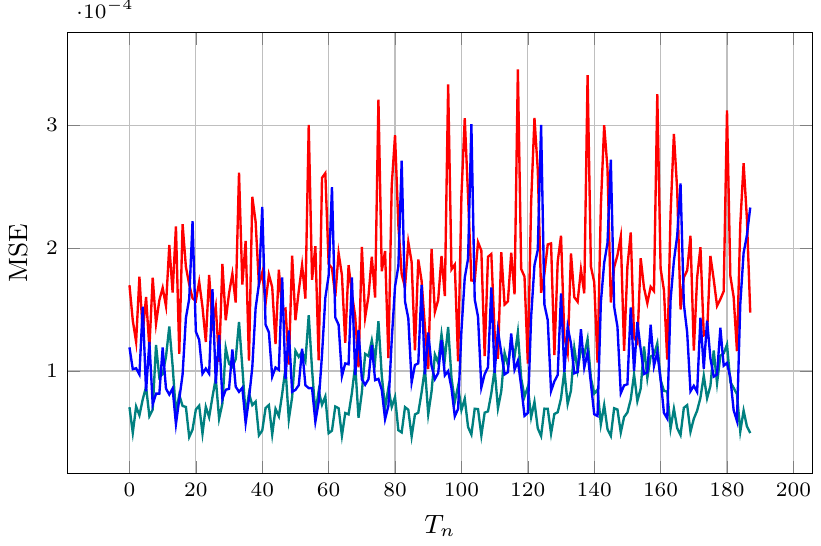}
	\caption{ }
\end{subfigure} \quad
\begin{subfigure}[b]{.45\linewidth}
	\centering
         \includegraphics{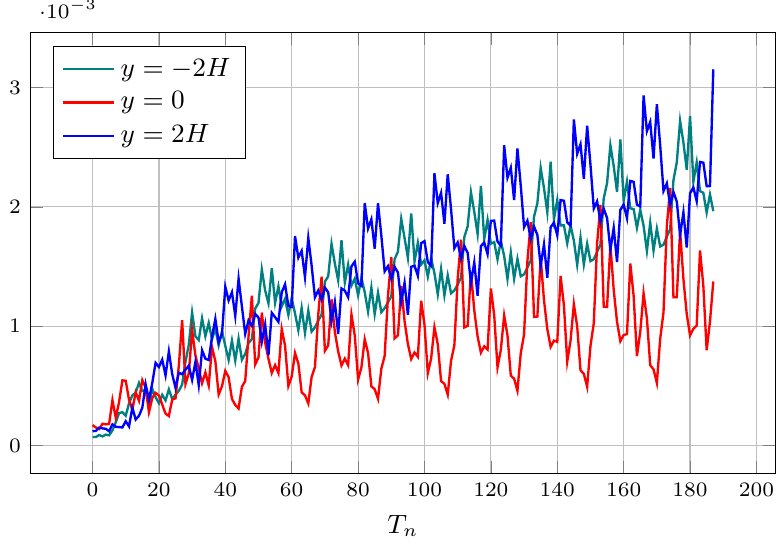}
	\caption{}
\end{subfigure}

\begin{subfigure}[b]{.45\linewidth}
	\centering
         \includegraphics{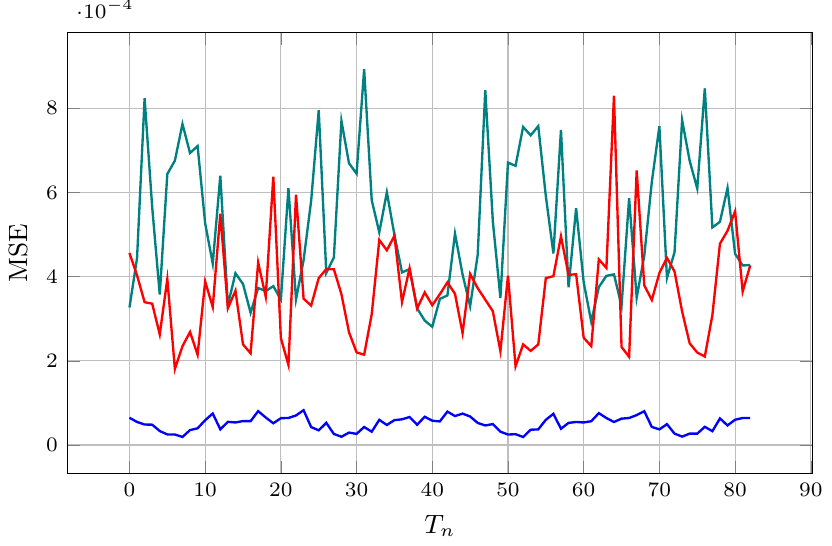}
	\caption{ }
\end{subfigure} \quad
\begin{subfigure}[b]{.45\linewidth}
	\centering
         \includegraphics{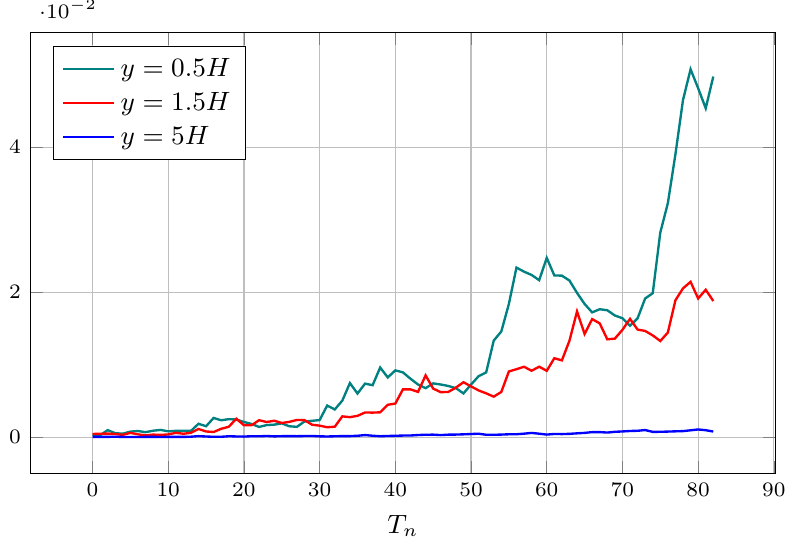}
	\caption{}
	\label{fig:tDevCase2}
\end{subfigure}

 \caption{\textbf{Mean squared error propagation} for velocity magnitude with respect to the reference values. The top row shows the evolution for case 1, and the bottom row shows the evolution for case 2. On the left is the temporal evolution of \textit{a priori} mean squared error, while on the right is the temporal evolution of \textit{a posteriori} mean squared error. The values are shown for locations along the Y-axis.} 
\label{fig:timeDistance_velMag_aY}

\end{figure}


Later, the temporal evolution of prediction error against reference by computing relative mean-squared errors of velocity magnitude for both cases is investigated. These errors are measured along the locations mentioned earlier. Figure \ref{fig:timeDistance_velMag}(a) shows the evolution of error for \textit{a priori} predictions and figure \ref{fig:timeDistance_velMag}(b) shows \textit{a posteriori} predictions for case 1 measured at streamwise locations. As could be expected, the errors accumulate for long-term posterior predictions, leading to a clear distinction when compared to \textit{a priori} predictions. It is interesting to note that although magnitude increases over time, this evolution also follows the vortex/wake shedding cycles denoting that the trained model performs well for long-term \textit{a posteriori} predictions. A similar trend is observed for case 2 as shown in figure \ref{fig:timeDistance_velMag}(c) and figure \ref{fig:timeDistance_velMag}(d), although here the magnitude of accumulated errors is higher than that of case 1. As shown in figure \ref{fig:timeDistance_velMag_aY}, a similar trend is observed when measurements of errors were performed at cross-streamwise locations.

\begin{figure}
\centering
\begin{subfigure}[b]{.45\linewidth}
	\centering
             \includegraphics{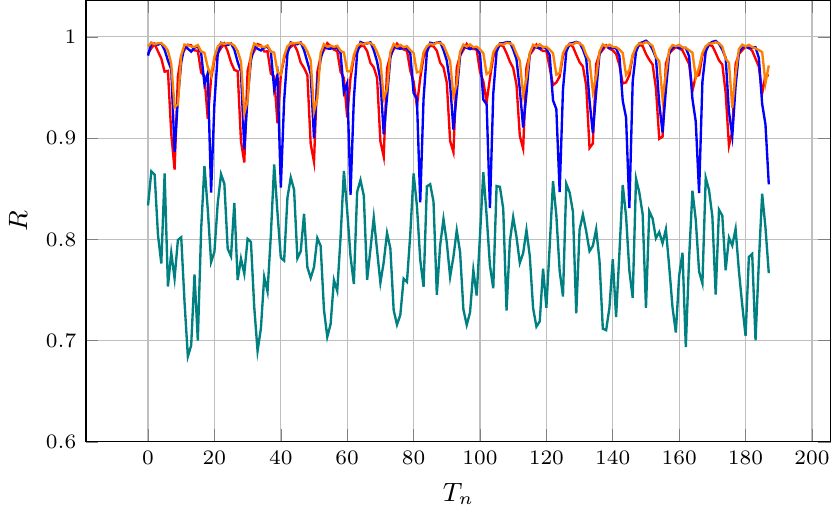}                    
	\caption{ }
\end{subfigure} \quad
\begin{subfigure}[b]{.45\linewidth}
	\centering
             \includegraphics{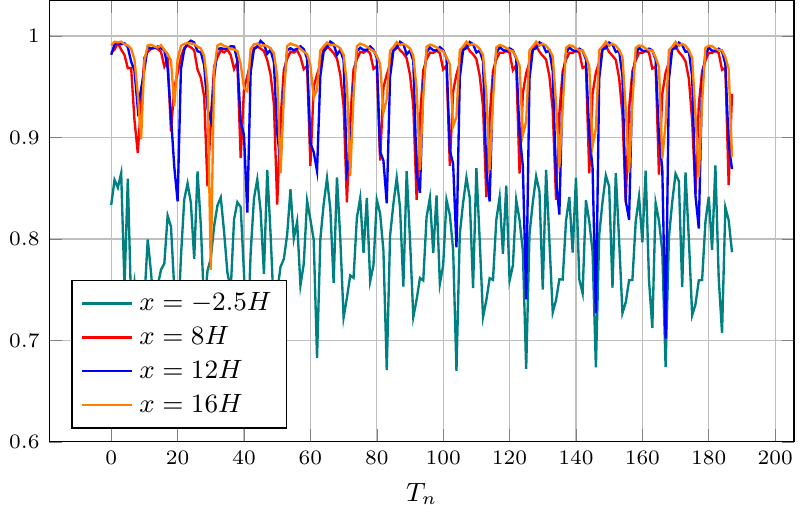}                    
	\caption{}
\end{subfigure}

\begin{subfigure}[b]{.45\linewidth}
	\centering
             \includegraphics{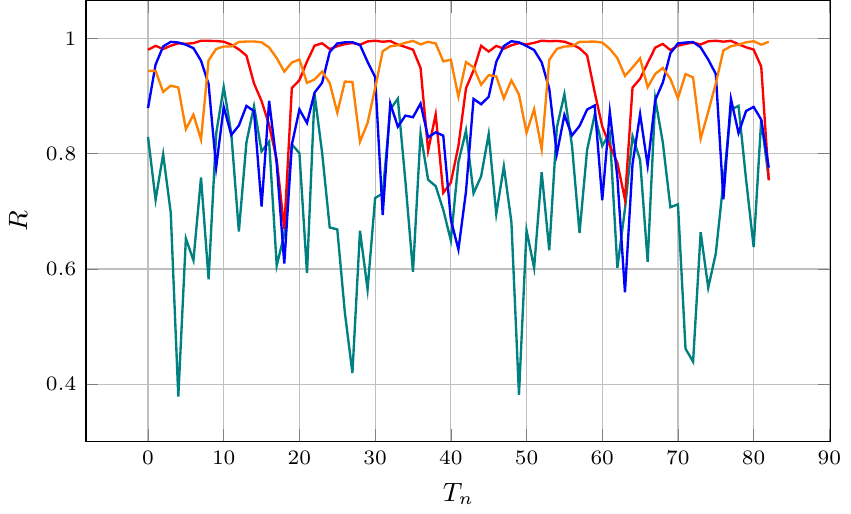}                    
	\caption{ }
\end{subfigure} \quad
\begin{subfigure}[b]{.45\linewidth}
	\centering
             \includegraphics{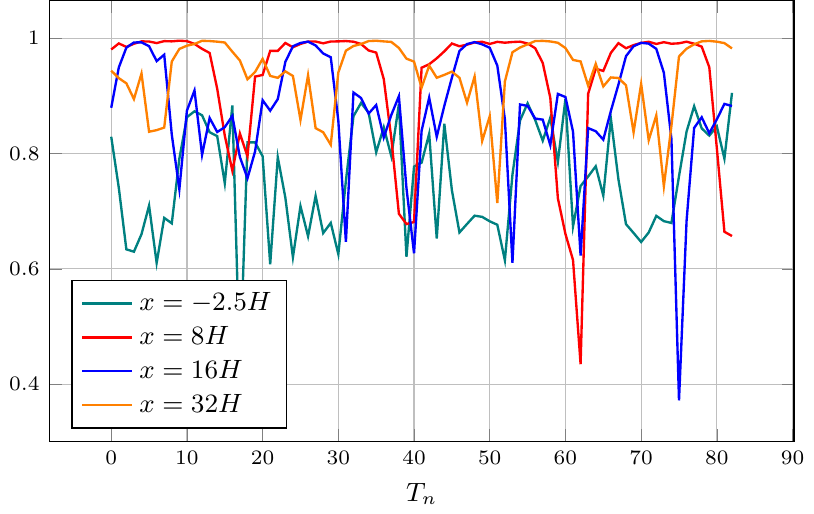}                    
	\caption{}
	\label{fig:tDevCase2}
\end{subfigure}

\caption{\textbf{Correlation propagation} for velocity magnitude with respect to the true values. The top row shows the evolution for case 1, and the bottom row shows the evolution for case 2. On the left is the temporal evolution of the Pearson product-moment correlation coefficient for \textit{a priori} values with reference to true values, while on the right are the $R$ values for \textit{a posteriori} values with reference to true values. The values are shown for locations along the X-axis.}
\label{fig:timePearsonR_velMag}

\end{figure}


\begin{figure}
\centering
\begin{subfigure}[b]{.45\linewidth}
	\centering
             \includegraphics{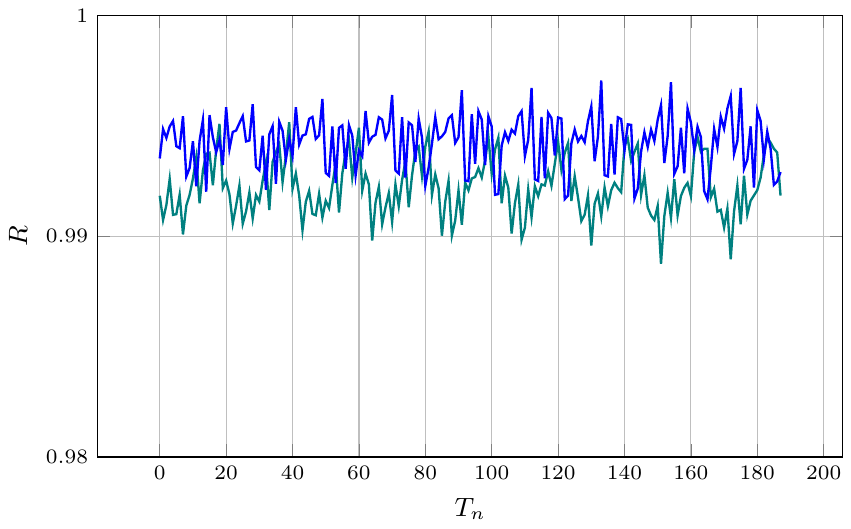}                    
	\caption{ }
\end{subfigure} \quad
\begin{subfigure}[b]{.45\linewidth}
	\centering
             \includegraphics{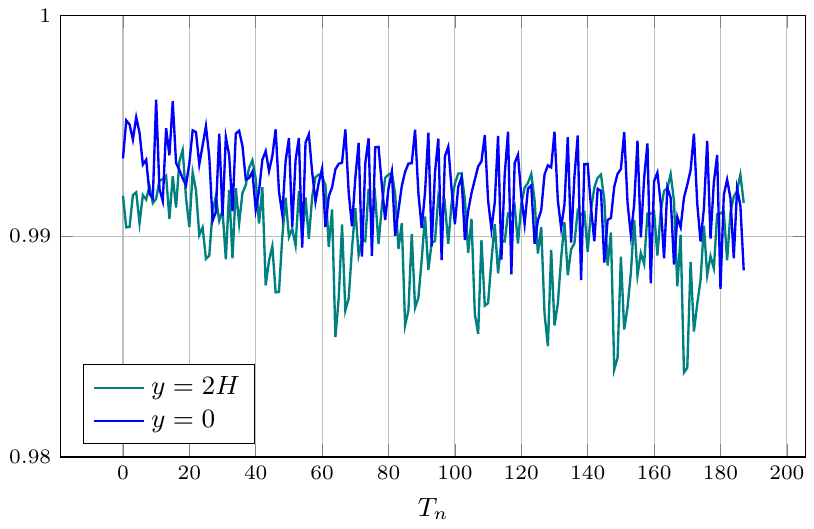}                    
	\caption{}
\end{subfigure}

\begin{subfigure}[b]{.45\linewidth}
	\centering
             \includegraphics{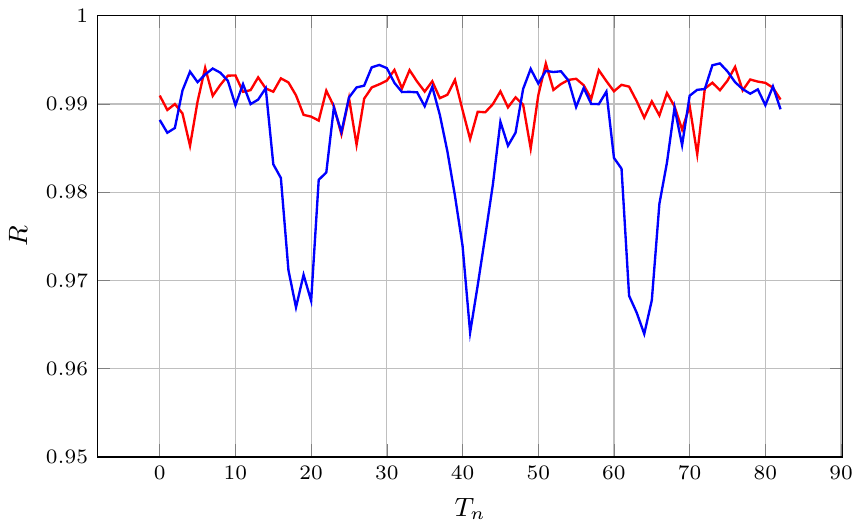}                    
	\caption{ }
\end{subfigure} \quad
\begin{subfigure}[b]{.45\linewidth}
	\centering
             \includegraphics{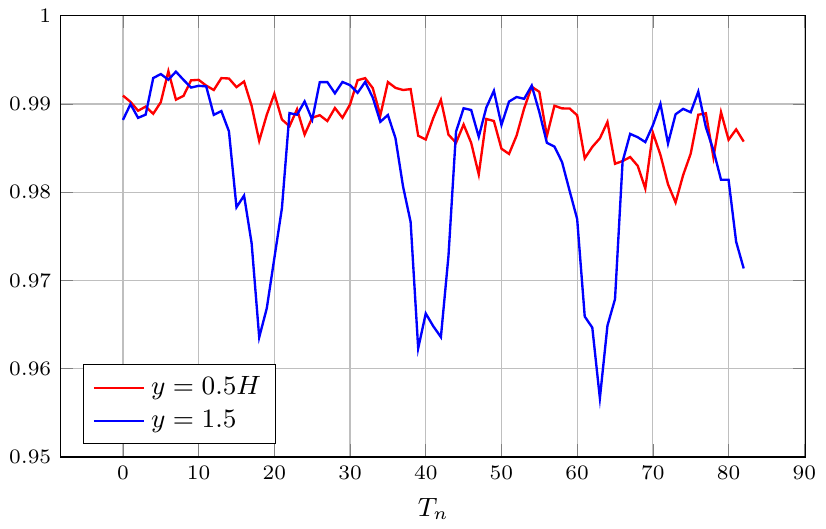}                    
	\caption{}
	\label{fig:tDevCase2}
\end{subfigure}

\caption{\textbf{Correlation propagation} for velocity magnitude with respect to the true values. The top row shows the evolution for case 1, and the bottom row shows the evolution for case 2. On the left is the temporal evolution of the Pearson product-moment correlation coefficient for \textit{a priori} values with reference to true values, while on the right are the $R$ values for \textit{a posteriori} values with reference to true values. The values are shown for locations along the Y-axis.}
\label{fig:timePearsonR_velMag_aY}

\end{figure}

\begin{figure}
\centering
\begin{subfigure}[b]{.45\linewidth}
	\centering
             \includegraphics{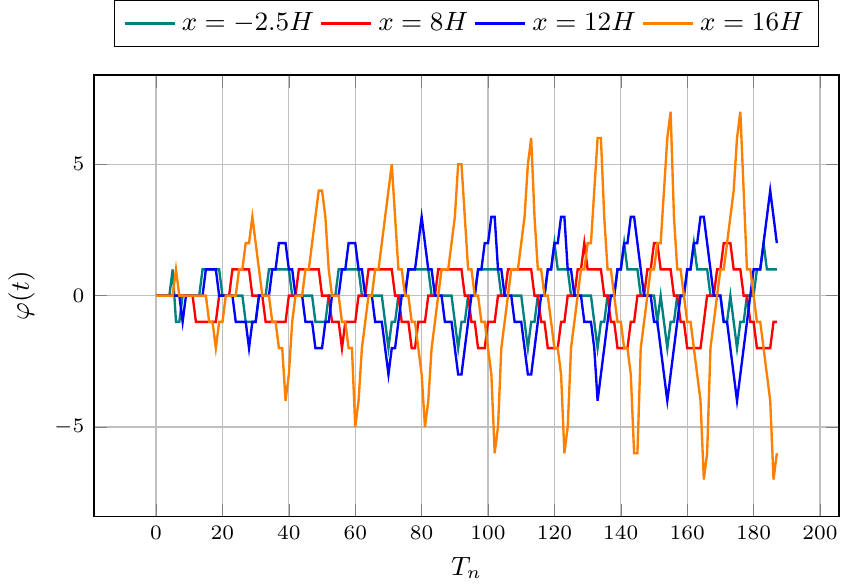}                    
	\caption{ }
\end{subfigure} \quad
\begin{subfigure}[b]{.45\linewidth}
	\centering
             \includegraphics{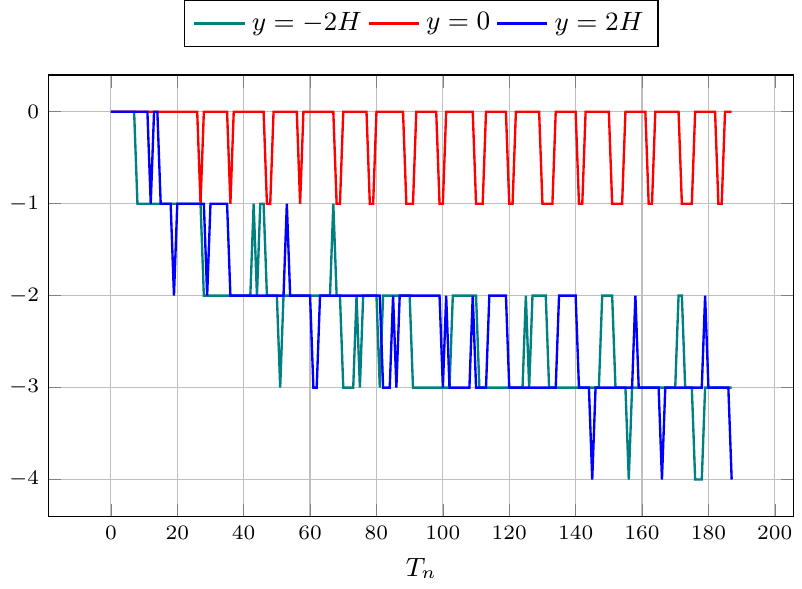}                    
	\caption{}
\end{subfigure}

\begin{subfigure}[b]{.45\linewidth}
	\centering
             \includegraphics{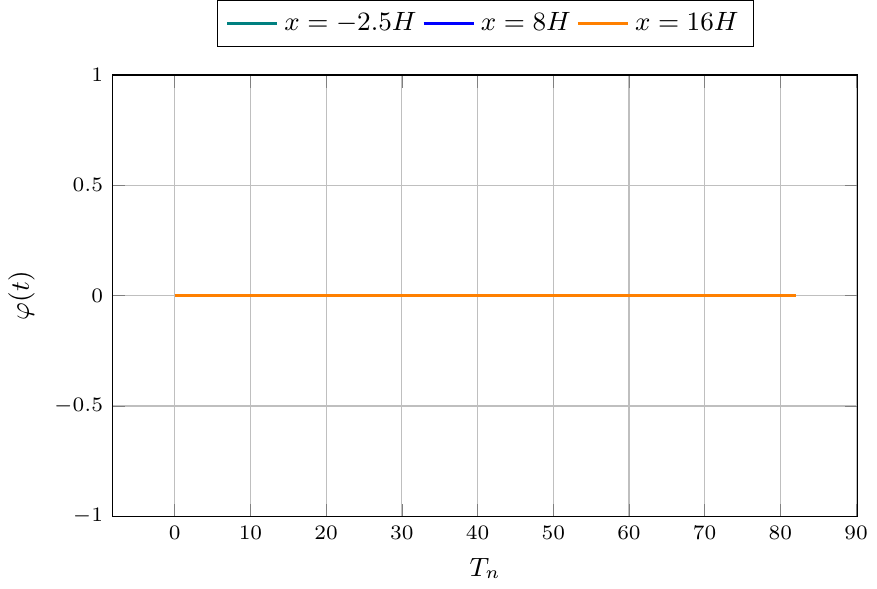}                    
	\caption{ }
\end{subfigure} \quad
\begin{subfigure}[b]{.45\linewidth}
	\centering
             \includegraphics{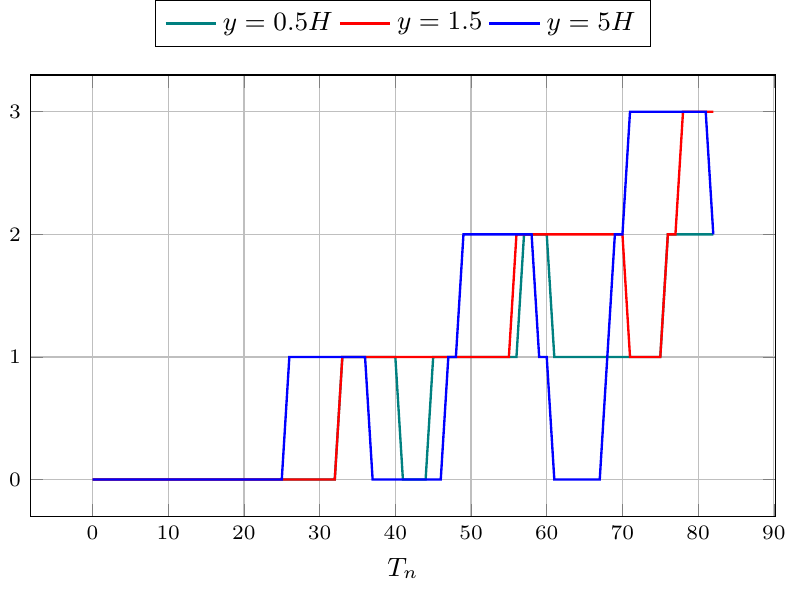}                    

	\caption{}
	\label{fig:tDevCase2}
\end{subfigure}

\caption{\textbf{Phase-shift evolution} for \textit{a posteriori} values of velocity magnitude with respect to the reference values. The top row shows the evolution for case 1, and the bottom row shows the evolution for case 2. On the left are temporal evolutions measured along with streamwise locations. While on the right are the evolutions measured along with cross-streamwise locations.}
\label{fig:timeShift_velMag}

\end{figure}

The \textit{a posteriori} predictions are observed to experience high noise conditions caused by error propagation. Hence, extracting the correlation between these two sets (predicted vs. reference) of temporal evolution is important, in particular to assess whether the heavy noise contributions are degrading correlation values.  To do so, the Pearson product-moment correlation coefficient $R_{xy}$ of $n$ pairs of time series data ${  \left\{(x_{1},y_{1}),\ldots ,(x_{n},y_{n})\right\}}$ is computed:

\begin{equation} 
R_{xy} = \frac{\sum ^n _{i=1}(x_i - \bar{x})(y_i - \bar{y})} {\sqrt{\sum ^n _{i=1}(x_i - \bar{x})^2} \sqrt{\sum ^n _{i=1}(y_i - \bar{y})^2}},
\end{equation} 

\noindent where $n$ is sample size, $x_i$, $y_i$  are the individual sample points indexed with $i$, and $ {\bar {x}}={\frac {1}{n}}\sum _{i=1}^{n}x_{i} $ is the sample mean (analogously for $\bar {y}$). In simple terms, $R_{xy}$ is the covariance of the two variables divided by the product of their standard deviations. For our measurements, the two variables are simply the predicted and reference snapshots at the same instants, and computation is performed for both the \textit{a priori} and \textit{a posteriori} predictions. Figure \ref{fig:timePearsonR_velMag}(b) and \ref{fig:timePearsonR_velMag}(d) show a gradual decrease in the correlation coefficient for the \textit{a posteriori} predictions for case 1 and case 2 respectively. A steeper degradation of correlation is observed in the measurements at cross-streamwise locations as shown in figures \ref{fig:timePearsonR_velMag_aY}(b) and \ref{fig:timePearsonR_velMag_aY}(d) for both cases, while that of the \textit{a posteriori} predictions remains stable. Since a clear trend is observed in degrading correlations for \textit{a posteriori} predictions, the phase-shift $\varphi(t)$ were measured for the temporal evolution of velocity magnitude predictions against the reference. Measurements were done along the similar spatial directions as mentioned before, the results of which are shown in figure \ref{fig:timeShift_velMag}. The value $\varphi(t) < 0$ denotes that predictions are shifted by that value before the reference, and the $\varphi(t)> 0$ denotes predictions shifted after the reference. For case 1, it is interesting to note that the phase shift goes on increasing in magnitude as time evolves, indicating the model's stability for long-term predictions. However, any clear trend for case 2 when measured at streamwise locations was not observed.  


\begin{figure}
\centering
\begin{subfigure}[b]{.3\linewidth}
	\centering
             \includegraphics{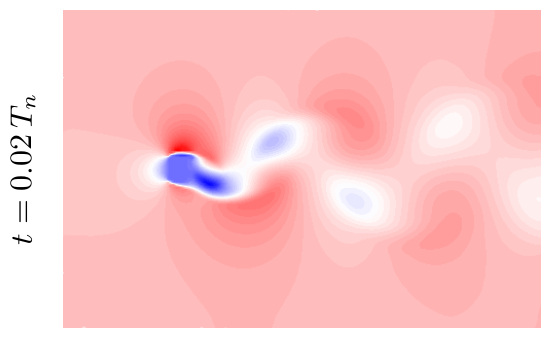}                    
	\caption*{}
\end{subfigure} \quad
\begin{subfigure}[b]{.3\linewidth}
	\centering
             \includegraphics{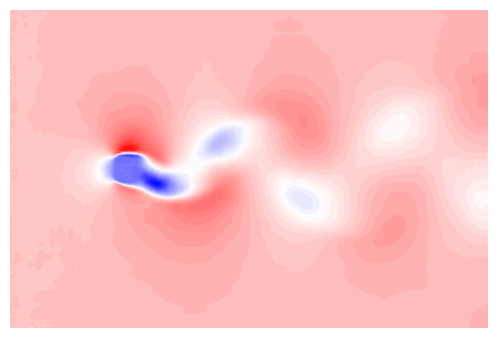}                    
	\caption*{}
\end{subfigure} \quad
\begin{subfigure}[b]{.3\linewidth}
	\centering
             \includegraphics{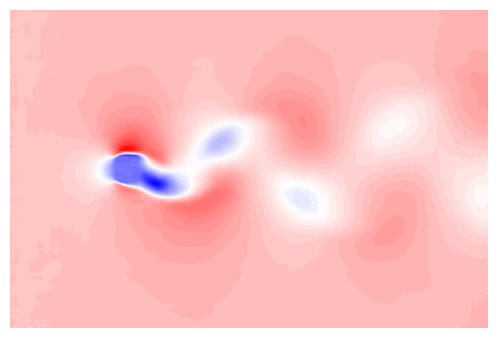}                    
	\caption*{}
\end{subfigure}

\begin{subfigure}[b]{.3\linewidth}
	\centering
             \includegraphics{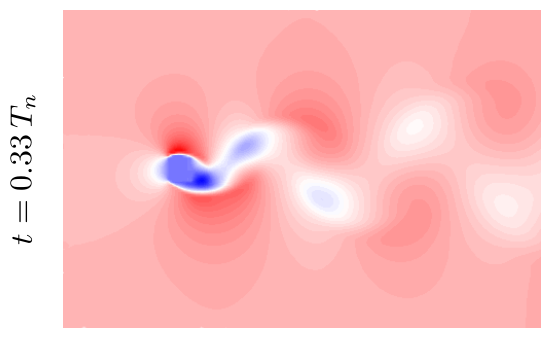}                    
	\caption*{}
\end{subfigure} \quad
\begin{subfigure}[b]{.3\linewidth}
	\centering
             \includegraphics{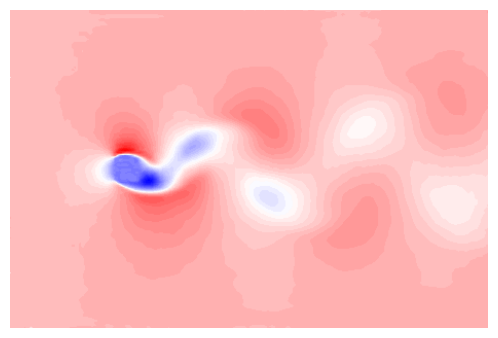}                    
	\caption*{}
\end{subfigure} \quad
\begin{subfigure}[b]{.3\linewidth}
	\centering
             \includegraphics{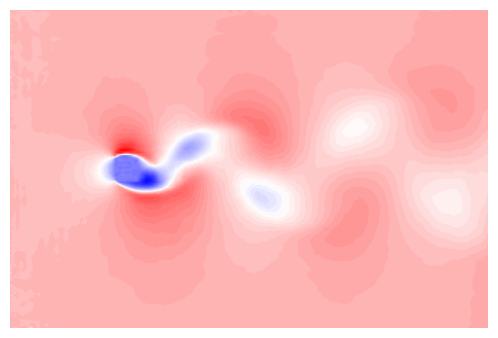}                    
	\caption*{}
\end{subfigure}

\begin{subfigure}[b]{.3\linewidth}
	\centering
             \includegraphics{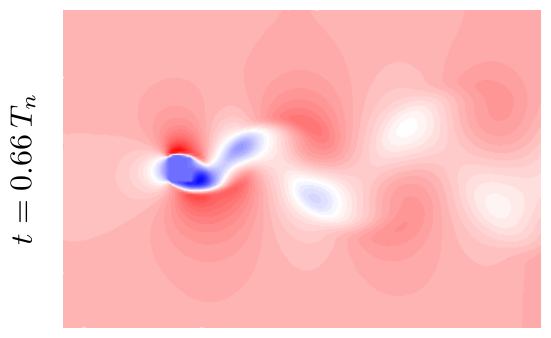}                    
	\caption*{}
\end{subfigure} \quad
\begin{subfigure}[b]{.3\linewidth}
	\centering
             \includegraphics{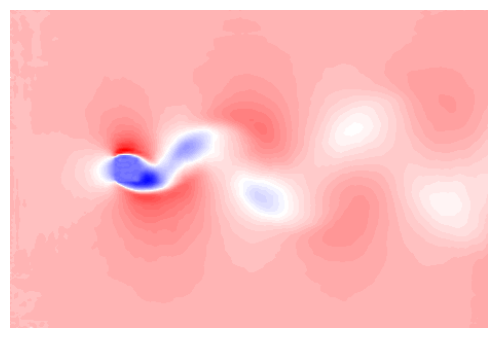}                    
	\caption*{}
\end{subfigure} \quad
\begin{subfigure}[b]{.3\linewidth}
	\centering
             \includegraphics{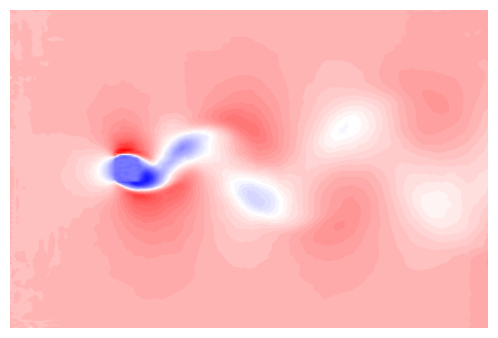}                    
	\caption*{}
\end{subfigure}

\begin{subfigure}[b]{.3\linewidth}
	\centering
             \includegraphics{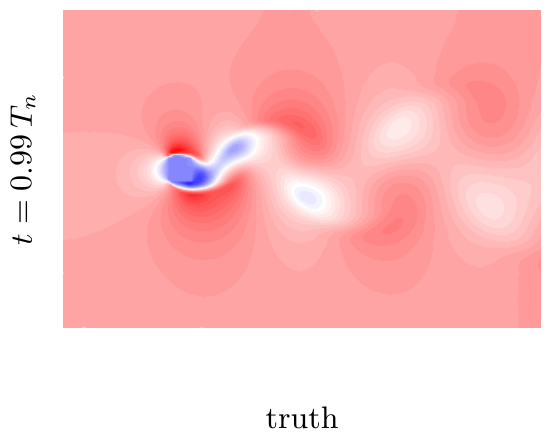}                    
	\caption*{}
\end{subfigure} \quad
\begin{subfigure}[b]{.3\linewidth}
	\centering
             \includegraphics{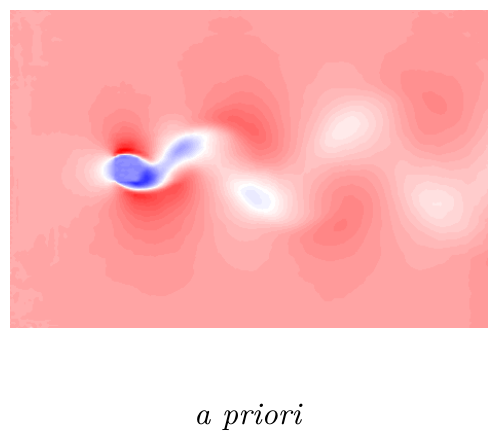}                    
	\caption*{}
\end{subfigure} \quad
\begin{subfigure}[b]{.3\linewidth}
	\centering
             \includegraphics{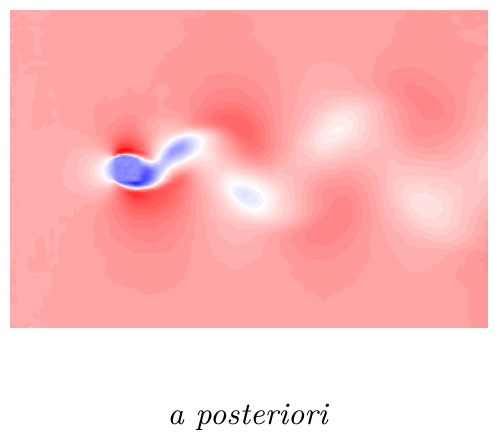}                    
	\caption*{}
\end{subfigure}

\vspace{-2em}%
\begin{subfigure}[t]{.9\linewidth}
	\centering
             \includegraphics{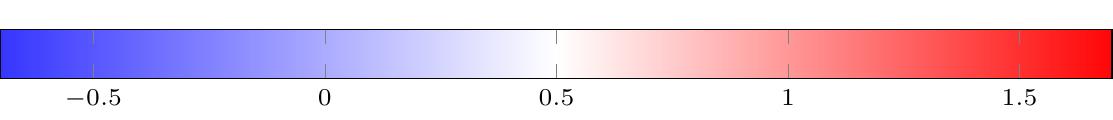}                    
	\caption*{}
\end{subfigure}

\caption{\textbf{Comparison of \textit{a priori} and \textit{a posteriori} prediction of streamwise velocity contours} against the reference showing the temporal evolution for case 1. }
\label{fig:cfdu0_comparison}
\end{figure}

\begin{figure}
\centering
\begin{subfigure}[b]{.3\linewidth}
	\centering
             \includegraphics{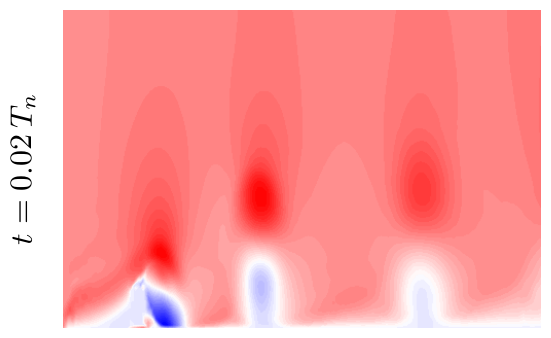}                    
	\caption*{}
\end{subfigure} \quad
\begin{subfigure}[b]{.3\linewidth}
	\centering
             \includegraphics{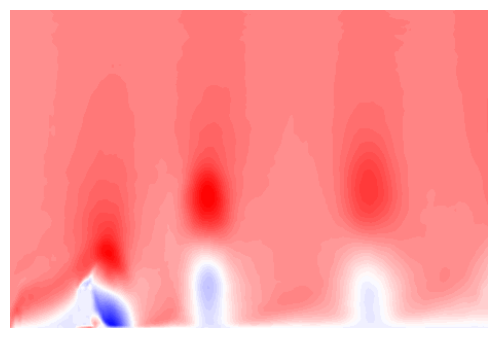}                    
	\caption*{}
\end{subfigure} \quad
\begin{subfigure}[b]{.3\linewidth}
	\centering
             \includegraphics{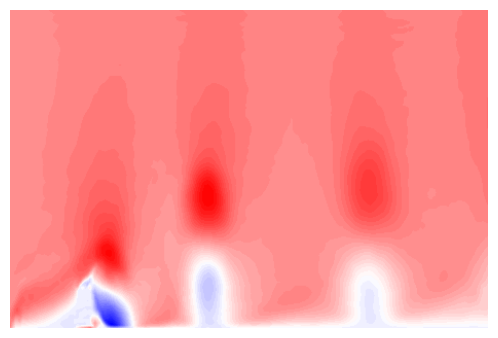}                    
	\caption*{}
\end{subfigure}

\begin{subfigure}[b]{.3\linewidth}
	\centering
             \includegraphics{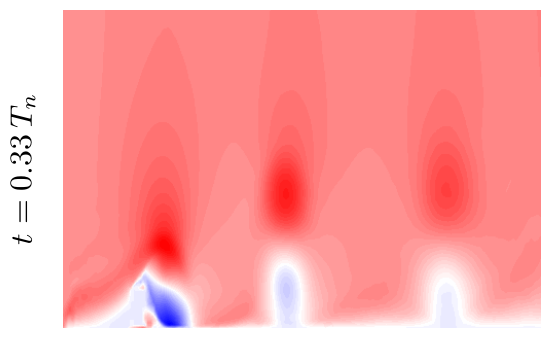}                    
	\caption*{}
\end{subfigure} \quad
\begin{subfigure}[b]{.3\linewidth}
	\centering
             \includegraphics{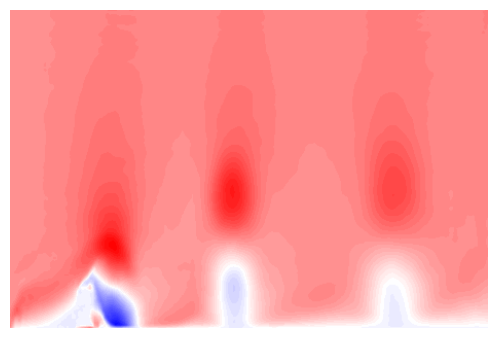}                    
	\caption*{}
\end{subfigure} \quad
\begin{subfigure}[b]{.3\linewidth}
	\centering
             \includegraphics{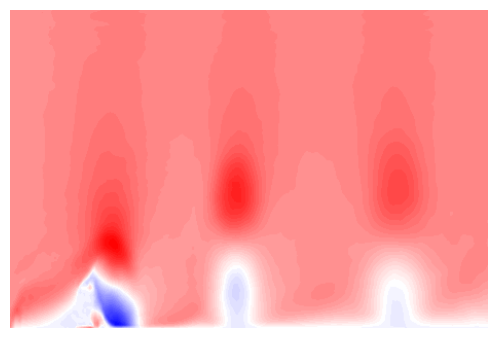}                    
	\caption*{}
\end{subfigure}

\begin{subfigure}[b]{.3\linewidth}
	\centering
             \includegraphics{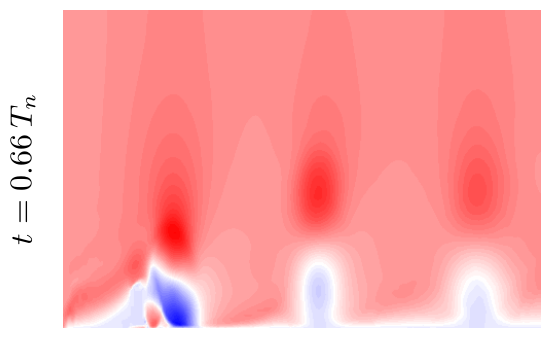}                    
	\caption*{}
\end{subfigure} \quad
\begin{subfigure}[b]{.3\linewidth}
	\centering
             \includegraphics{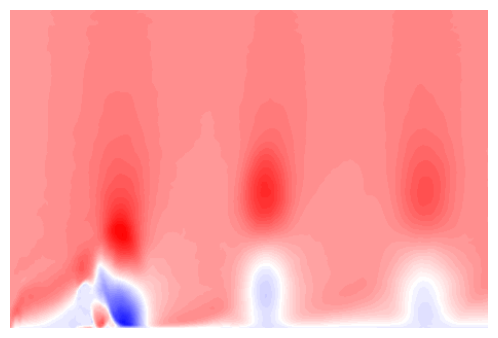}                    
	\caption*{}
\end{subfigure} \quad
\begin{subfigure}[b]{.3\linewidth}
	\centering
             \includegraphics{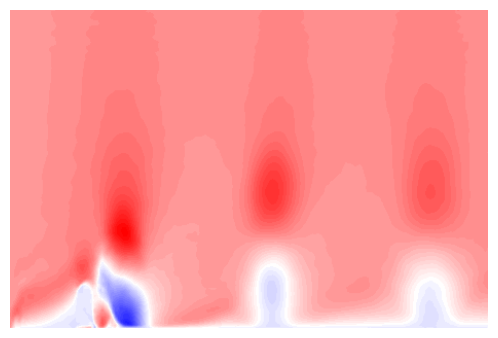}                    
	\caption*{}
\end{subfigure}

\begin{subfigure}[b]{.3\linewidth}
	\centering
             \includegraphics{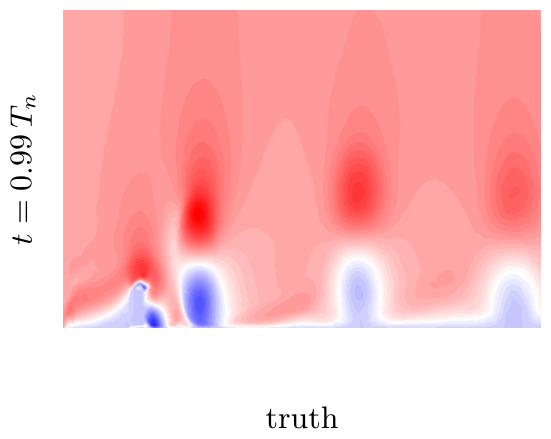}                    
	\caption*{}
\end{subfigure} \quad
\begin{subfigure}[b]{.3\linewidth}
	\centering
             \includegraphics{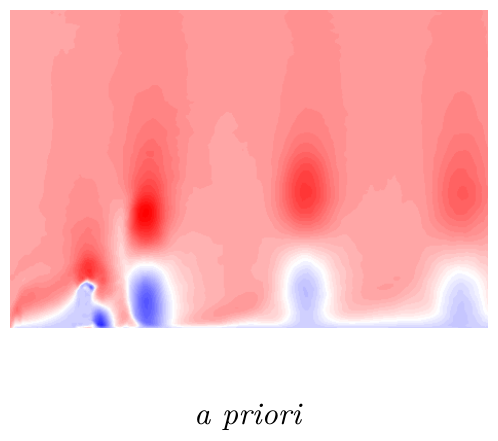}                    
	\caption*{}
\end{subfigure} \quad
\begin{subfigure}[b]{.3\linewidth}
	\centering
             \includegraphics{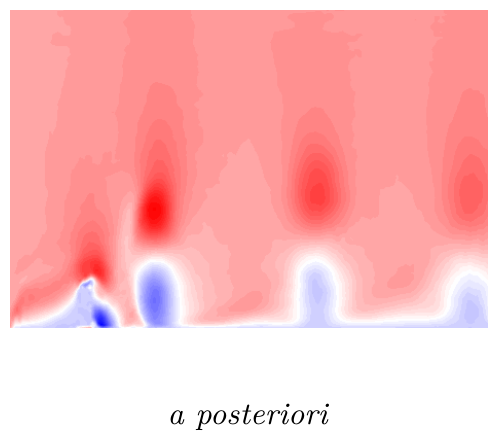}                    
	\caption*{}
\end{subfigure}

\vspace{-2em}%
\begin{subfigure}[t]{.9\linewidth}
	\centering
             \includegraphics{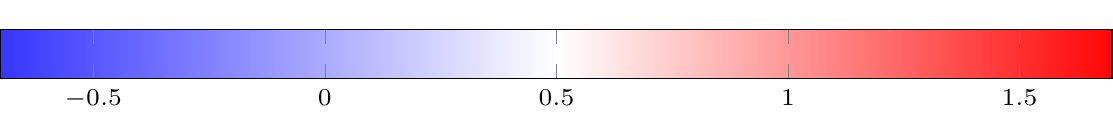}                    
	\caption*{}
\end{subfigure}

\caption{\textbf{Comparison of \textit{a priori} and \textit{a posteriori} prediction of streamwise velocity contours} against the truth reference showing the temporal evolution for case 2. }
\label{fig:cfdu0_comparisonC2}

\end{figure}

For a qualitative assessment of results, the contours of velocity components for both cases is compared.  Figure \ref{fig:cfdu0_comparison} shows the instantaneous snapshots of streamwise velocity contours for case 1 at $t=[ 2\%, 33\%, 66\% ]$ of total predicted snapshots as mentioned earlier, and similar instantaneous snapshots for case 2 are shown in figure \ref{fig:cfdu0_comparisonC2}. For both the cases, the \textit{a priori}, as well as \textit{a posteriori} predictions, show a fairly accurate agreement with the reference.

\section{Conclusions}

A convolutional encoder-decoder-based transformer model has been developed to auto-regressively train on spatio-temporal data of turbulent flows. The method of auto-regressive training works by predicting future fluid flow fields from the previously predicted fluid flow field to ensure long-term predictions without diverging. The model has been validated by demonstrating its applicability to turbulent wake flow past an obstacle and environmental flow past surface mounted obstacle. The work demonstrates a promising model and method for forecasting fluid flow fields where the training data is available. The proposed model trained in an autoregressive way shows significant agreements for a priori evaluations, whereas the posterior predictions show expected deviations after a considerable number of simulation steps. The spatio-temporal complexity of predictions is comparable to the target simulations of fully developed turbulence. The autoregressive training and prediction of \textit{a posteriori} states is the primary step towards the development of more complex data-driven turbulence models and simulations. It is shown that the self-attention transformers incorporated within the convolutional encoder-decoder can predict up to $200\Delta t$ time steps with relatively high accuracy, and the proposed data-driven deep learning model remains stable for multiple long time scales, promising a stable and physical deep learning predictive turbulence modeling candidate. Changes to loss function can be done to achieve even longer, stable, physically realistic results. Additional experiments are needed to demonstrate the model's ability on generalizing to local mesh regions as well as longer \textit{a posteriori} simulation steps.  Further investigations on a variety of industrial and academic cases could include training for flow Reynolds numbers, turbulence intensity, and other inlet parameters. Conclusions from this work would also provide valuable insights for the development of new deep learning methods and their deployment for turbulent flows on complex geometries in industrial problems. Deploying a trained model to assist a fluid solver is regarded as a future extension of the present work.

\section{Data Availability}
The data and code that support the findings of this study are openly available at \url{https://github.com/aakash30jan/Spatio-Temporal-Learning-of-Turbulent-Flows} .

\section{acknowledgments}
This work is supported by the Carnot M.I.N.E.S. Institute through the MINDS - Mines Initiative for Numerics and Data Science project.

\appendix
\section{Appendixes}

\begin{figure}
\centering
\begin{subfigure}[b]{.3\linewidth}
	\centering
             \includegraphics{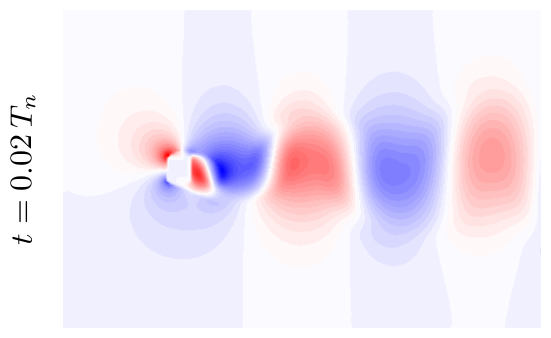}                    
	\caption*{}
\end{subfigure} \quad
\begin{subfigure}[b]{.3\linewidth}
	\centering
             \includegraphics{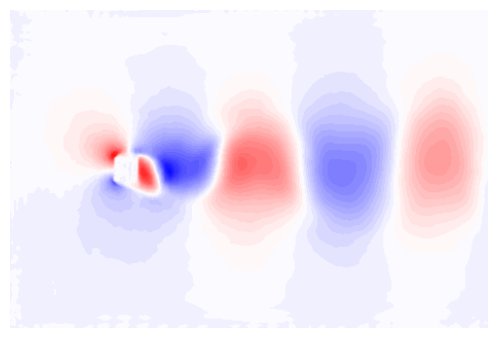}                    
	\caption*{}
\end{subfigure} \quad
\begin{subfigure}[b]{.3\linewidth}
	\centering
             \includegraphics{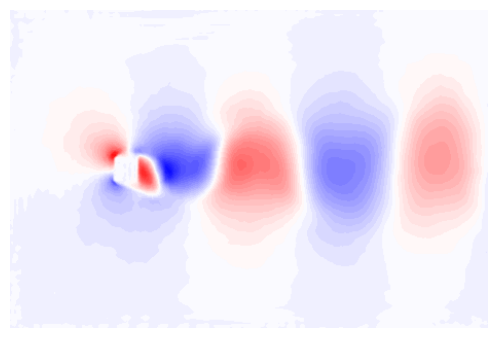}                    
	\caption*{}
\end{subfigure}

\begin{subfigure}[b]{.3\linewidth}
	\centering
             \includegraphics{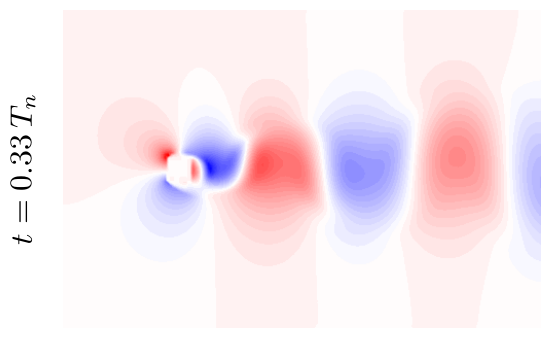}                    
	\caption*{}
\end{subfigure} \quad
\begin{subfigure}[b]{.3\linewidth}
	\centering
             \includegraphics{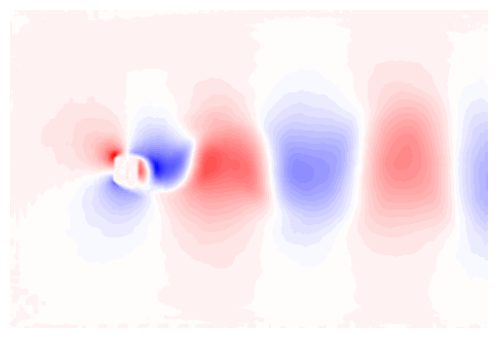}                    
	\caption*{}
\end{subfigure} \quad
\begin{subfigure}[b]{.3\linewidth}
	\centering
             \includegraphics{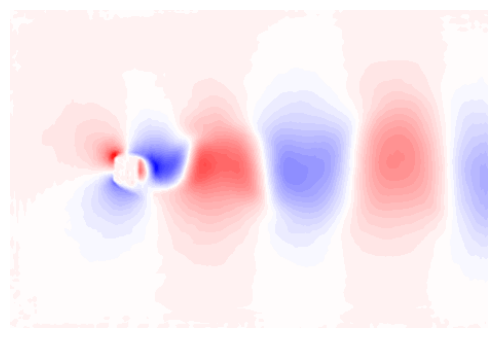}                    
	\caption*{}
\end{subfigure}

\begin{subfigure}[b]{.3\linewidth}
	\centering
             \includegraphics{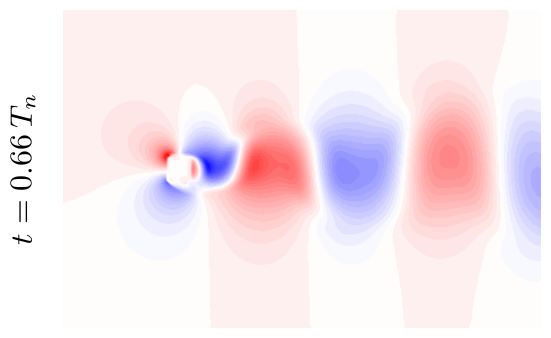}                    
	\caption*{}
\end{subfigure} \quad
\begin{subfigure}[b]{.3\linewidth}
	\centering
             \includegraphics{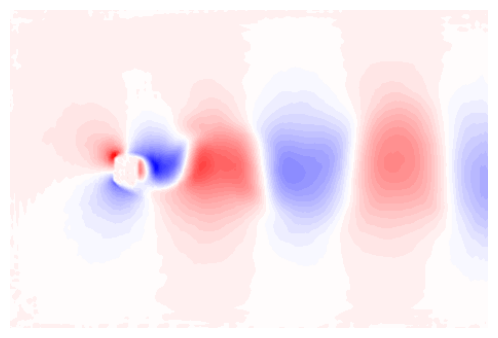}                    
	\caption*{}
\end{subfigure} \quad
\begin{subfigure}[b]{.3\linewidth}
	\centering
             \includegraphics{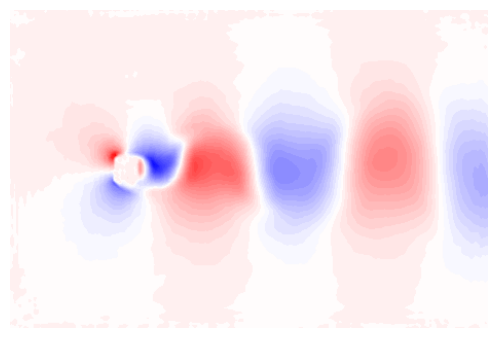}                    
	\caption*{}
\end{subfigure}

\begin{subfigure}[b]{.3\linewidth}
	\centering
             \includegraphics{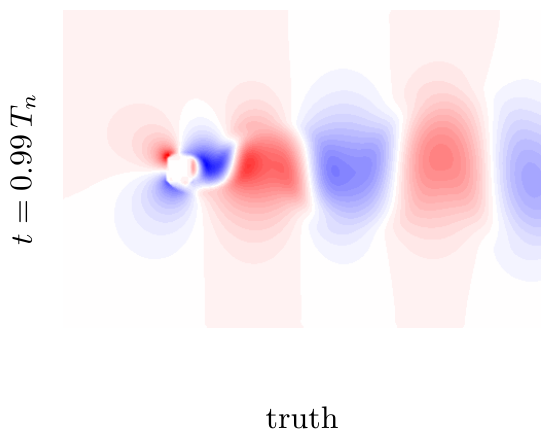}                    
	\caption*{}
\end{subfigure} \quad
\begin{subfigure}[b]{.3\linewidth}
	\centering
             \includegraphics{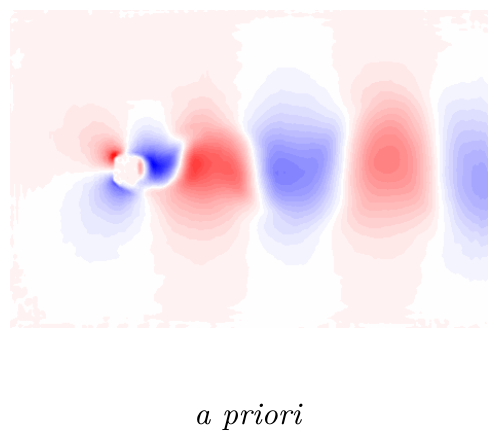}                    
	\caption*{}
\end{subfigure} \quad
\begin{subfigure}[b]{.3\linewidth}
	\centering
             \includegraphics{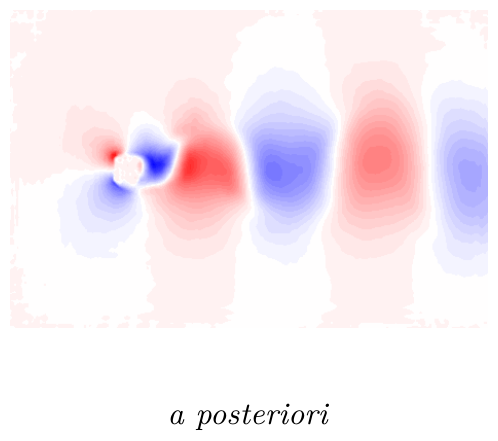}                    
	\caption*{}
\end{subfigure}

\vspace{-2em}%
\begin{subfigure}[t]{.9\linewidth}
	\centering
             \includegraphics{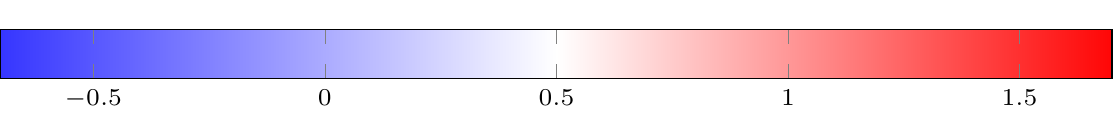}                    
	\caption*{}
\end{subfigure}

\caption{\textbf{Comparison of \textit{a priori} and \textit{a posteriori} prediction of cross streamwise velocity contours} against the reference showing the temporal evolution for case 1. }
\label{fig:cfdu1_comparison}
\end{figure}

\begin{figure}
\centering
\begin{subfigure}[b]{.3\linewidth}
	\centering
             \includegraphics{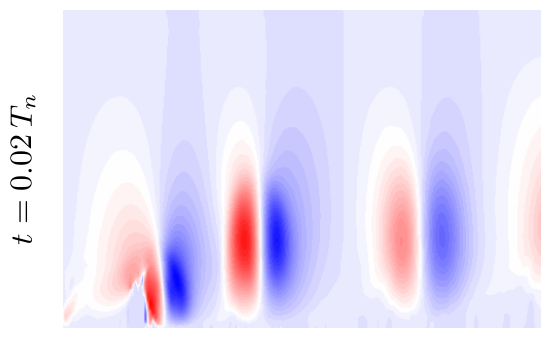}                    
	\caption*{}
\end{subfigure} \quad
\begin{subfigure}[b]{.3\linewidth}
	\centering
             \includegraphics{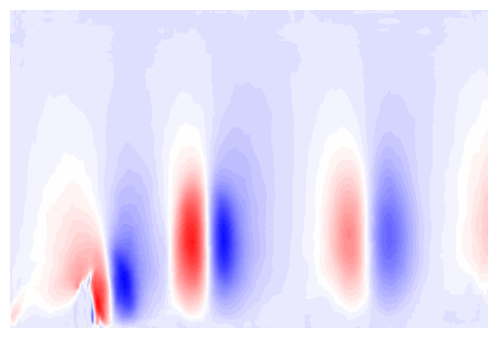}                    
	\caption*{}
\end{subfigure} \quad
\begin{subfigure}[b]{.3\linewidth}
	\centering
             \includegraphics{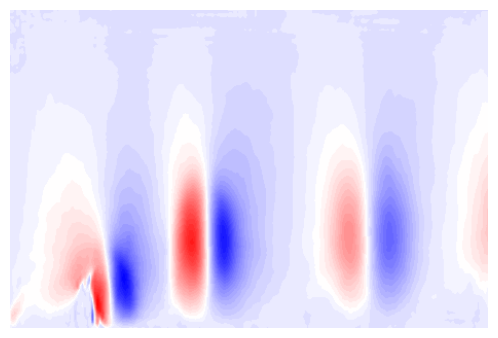}                    
	\caption*{}
\end{subfigure}

\begin{subfigure}[b]{.3\linewidth}
	\centering
             \includegraphics{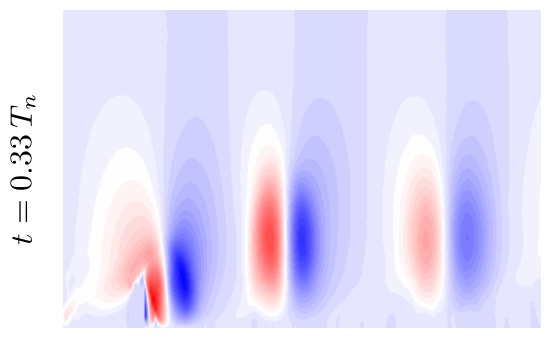}                    
	\caption*{}
\end{subfigure} \quad
\begin{subfigure}[b]{.3\linewidth}
	\centering
             \includegraphics{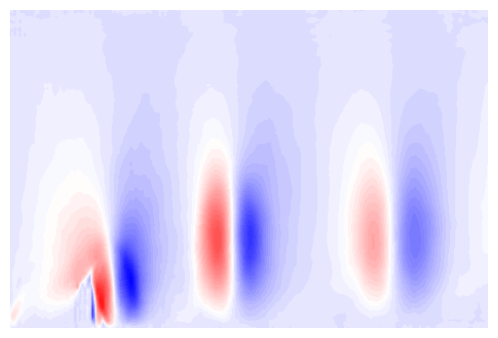}                    
	\caption*{}
\end{subfigure} \quad
\begin{subfigure}[b]{.3\linewidth}
	\centering
             \includegraphics{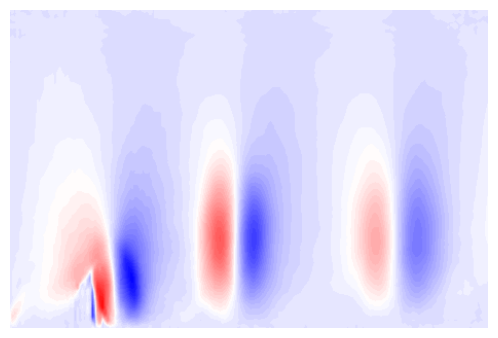}                    
	\caption*{}
\end{subfigure}

\begin{subfigure}[b]{.3\linewidth}
	\centering
             \includegraphics{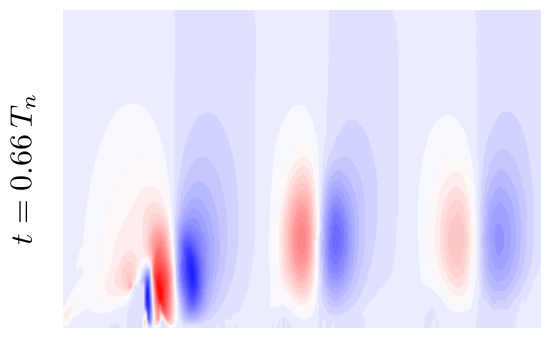}                    
	\caption*{}
\end{subfigure} \quad
\begin{subfigure}[b]{.3\linewidth}
	\centering
             \includegraphics{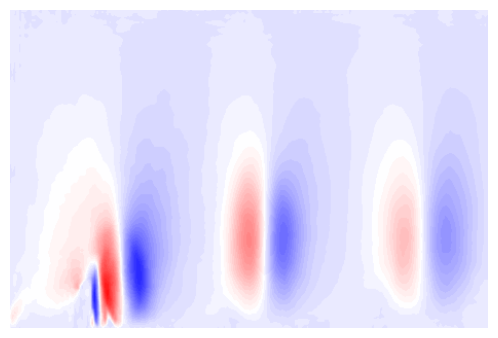}                    
	\caption*{}
\end{subfigure} \quad
\begin{subfigure}[b]{.3\linewidth}
	\centering
             \includegraphics{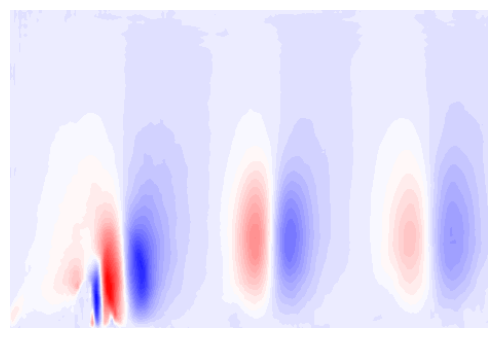}                    
	\caption*{}
\end{subfigure}

\begin{subfigure}[b]{.3\linewidth}
	\centering
             \includegraphics{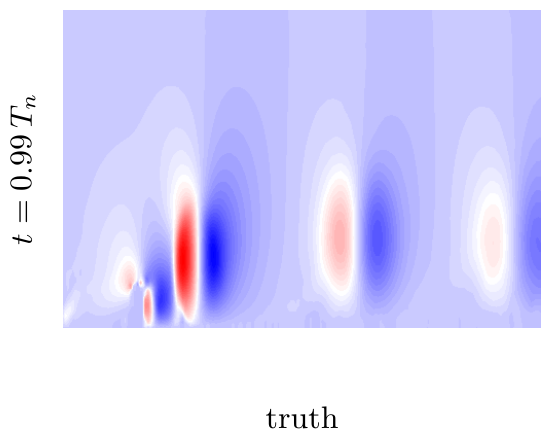}                    
	\caption*{}
\end{subfigure} \quad
\begin{subfigure}[b]{.3\linewidth}
	\centering
             \includegraphics{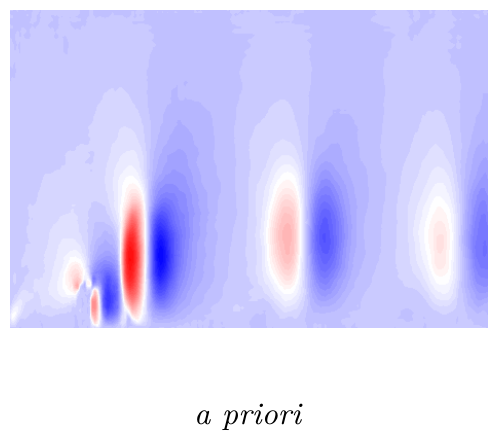}                    
	\caption*{}
\end{subfigure} \quad
\begin{subfigure}[b]{.3\linewidth}
	\centering
             \includegraphics{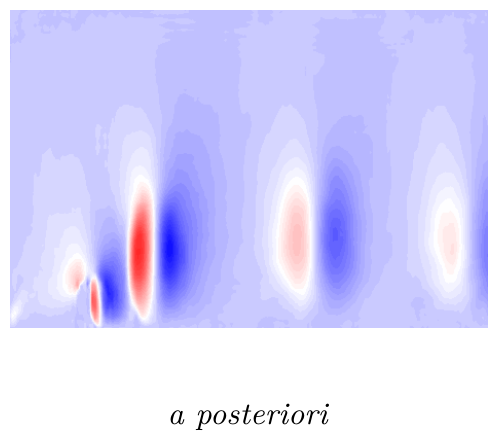}                    
	\caption*{}
\end{subfigure}

\vspace{-2em}%
\begin{subfigure}[t]{.9\linewidth}
	\centering
             \includegraphics{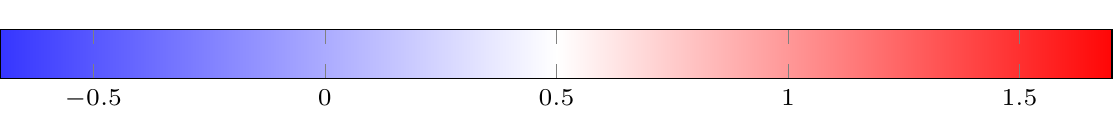}                    
	\caption*{}
\end{subfigure}

\caption{\textbf{Comparison of \textit{a priori} and \textit{a posteriori} prediction of cross streamwise velocity contours} against the truth reference showing the temporal evolution for case 2. }
\label{fig:cfdu1_comparisonC2}

\end{figure}

\begin{figure} 
\centering

\begin{subfigure}[t]{.9\linewidth}
	\centering
     \includegraphics{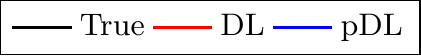}
	\caption*{}
\end{subfigure}
\vspace{-1em}%

\begin{subfigure}[b]{\linewidth}
\centering
\begin{subfigure}[b]{.4\linewidth}
     \includegraphics{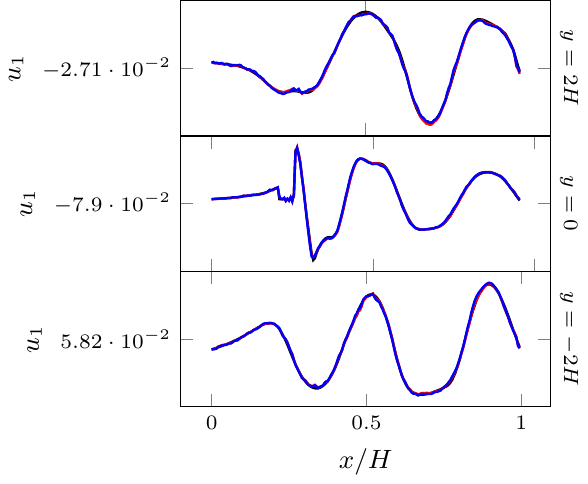}
\end{subfigure} \quad
\begin{subfigure}[b]{.4\linewidth}
      \includegraphics{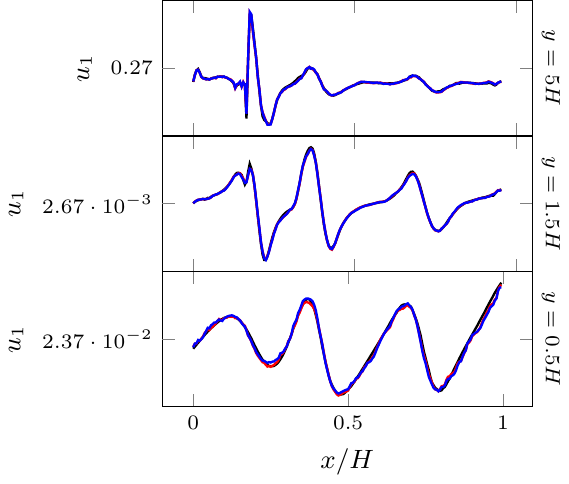}
\end{subfigure}  
\caption{Time $t= 0.02\, T_n$ }
\end{subfigure} 

\medskip
\medskip

\centering
\begin{subfigure}[b]{\linewidth}
\centering
\begin{subfigure}[b]{.4\linewidth}
     \includegraphics{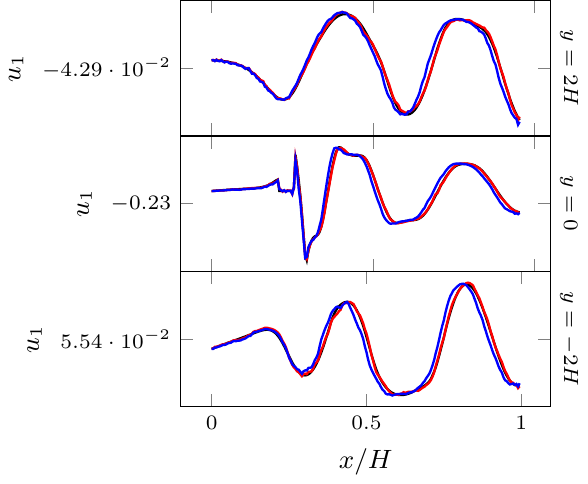}
\end{subfigure} \quad
\begin{subfigure}[b]{.4\linewidth}
      \includegraphics{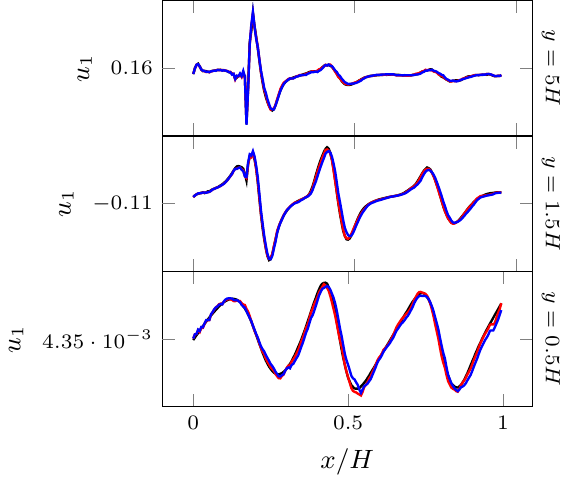}
\end{subfigure} 
\caption{Time $t= 0.33\, T_n$ }
\end{subfigure} 

\medskip
\medskip

\centering
\begin{subfigure}[b]{\linewidth}
\centering
\begin{subfigure}[b]{.4\linewidth}
      \includegraphics{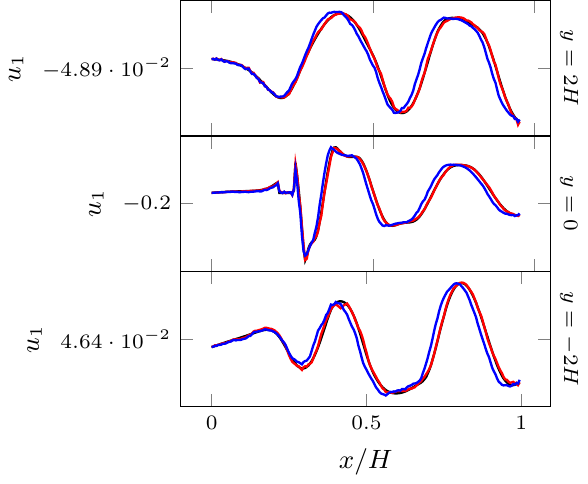}
\end{subfigure} \quad
\begin{subfigure}[b]{.4\linewidth}
         \includegraphics{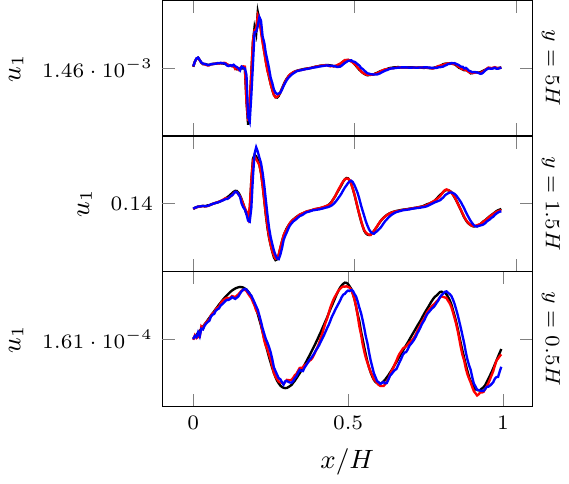}
\end{subfigure} 
\caption{Time $t= 0.66\, T_n$ }
\end{subfigure} 

\caption{\textbf{Comparative predictions of cross streamwise velocity components ($u_1$) sampled along y-axis for case 1 (left) and case 2 (right).} Figures from top to bottom denote the predictions at increasing times, \textit{i.e.} the top row contains instantaneous preditctions at $t= 0.02\, T_n$,  the middle row at $t= 0.33\, T_n$, and the bottom row shows the predictions at $t= 0.66\, T_n$ . }
\label{fig:comparison_pripost_TV_aY}
\end{figure}

\begin{figure} 
\centering

\begin{subfigure}[t]{.9\linewidth}
	\centering
     \includegraphics{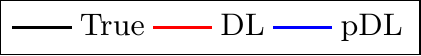}
	\caption*{}
\end{subfigure}
\vspace{-1em}%

\begin{subfigure}[b]{\linewidth}
\centering
\begin{subfigure}[b]{.4\linewidth}
     \includegraphics{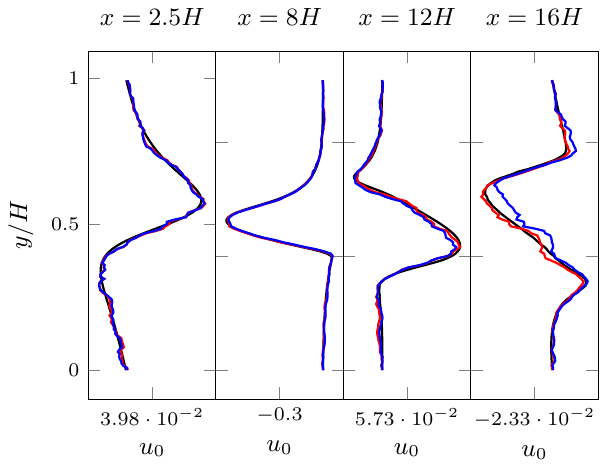}
\end{subfigure} \quad
\begin{subfigure}[b]{.4\linewidth}
      \includegraphics{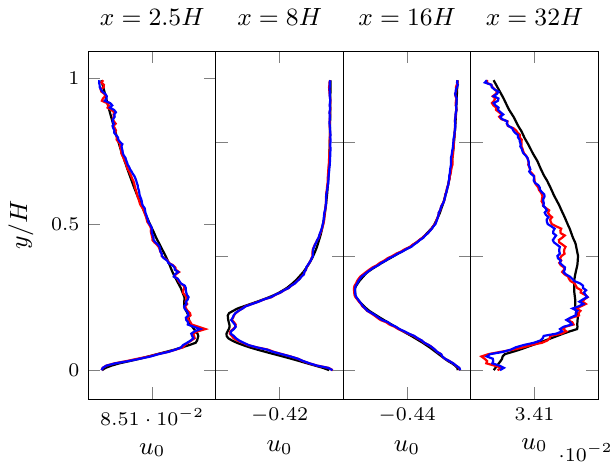}
\end{subfigure}  
\caption{Time $t= 0.02\, T_n$ }
\end{subfigure} 

\medskip
\medskip

\centering
\begin{subfigure}[b]{\linewidth}
\centering
\begin{subfigure}[b]{.4\linewidth}
     \includegraphics{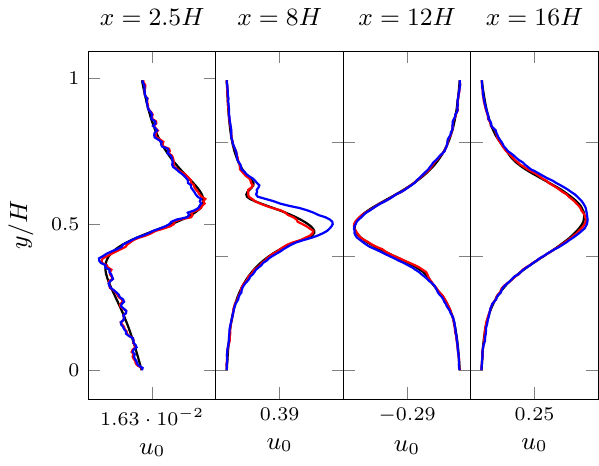}
\end{subfigure} \quad
\begin{subfigure}[b]{.4\linewidth}
      \includegraphics{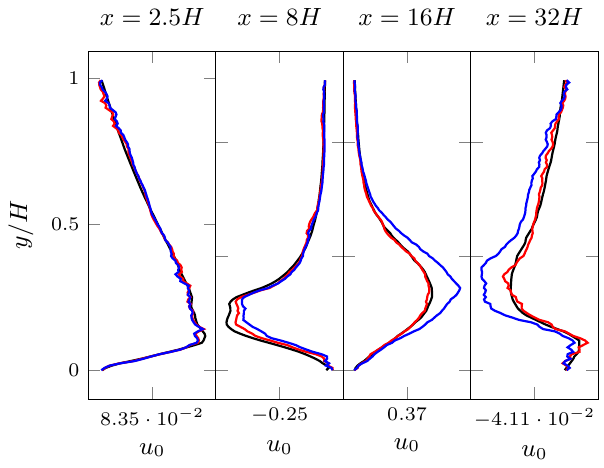}
\end{subfigure} 
\caption{Time $t= 0.33\, T_n$ }
\end{subfigure} 

\medskip
\medskip

\centering
\begin{subfigure}[b]{\linewidth}
\centering
\begin{subfigure}[b]{.4\linewidth}
      \includegraphics{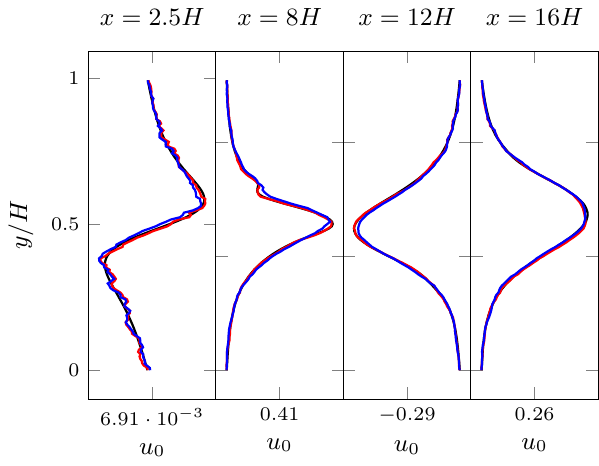}
\end{subfigure} \quad
\begin{subfigure}[b]{.4\linewidth}
         \includegraphics{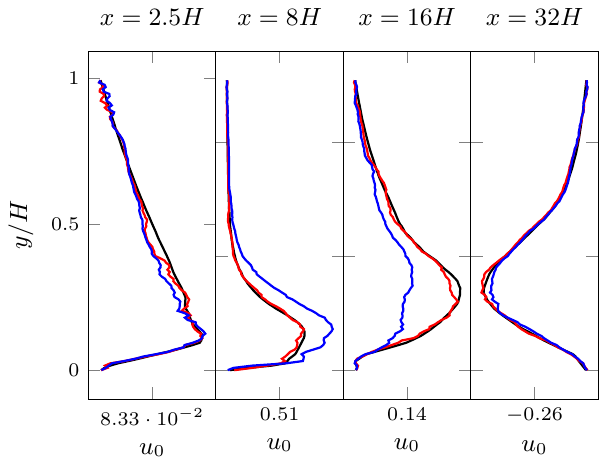}
\end{subfigure} 
\caption{Time $t= 0.66\, T_n$ }
\end{subfigure} 

\caption{\textbf{Comparative predictions of cross streamwise velocity components ($u_1$) for case 1 (left) and case 2 (right).} Figures from top to bottom, denote the predictions at increasing times, \textit{i.e.} the top row contains instantaneous predictions at $t= 0.02\, T_n$,  the middle row at $t= 0.33\, T_n$, and the bottom row shows the predictions at $t= 0.66\, T_n$.}
\label{fig:comparison_pripost_TV}
\end{figure}

\begin{figure} 
\centering

\begin{subfigure}[t]{.9\linewidth}
	\centering
     \includegraphics{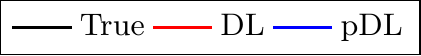}
	\caption*{}
\end{subfigure}
\vspace{-1em}%

\begin{subfigure}[b]{\linewidth}
\centering
\begin{subfigure}[b]{.4\linewidth}
     \includegraphics{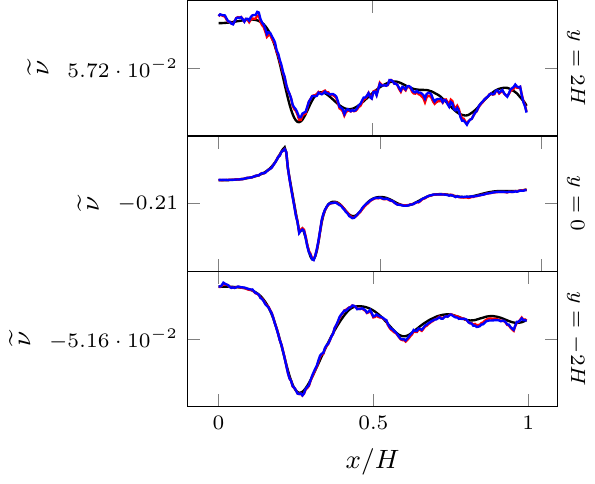}
\end{subfigure} \quad
\begin{subfigure}[b]{.4\linewidth}
      \includegraphics{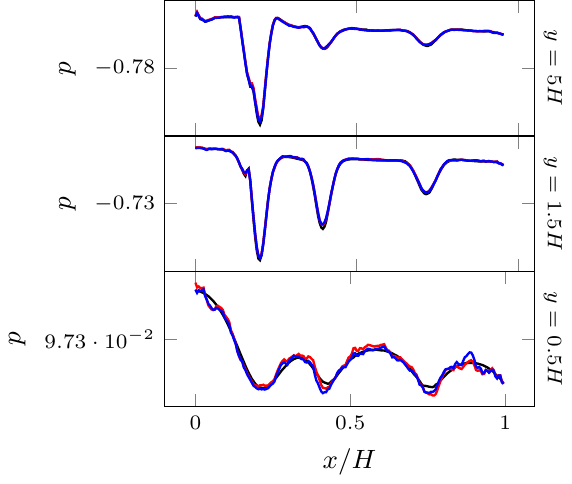}
\end{subfigure}  
\caption{Time $t= 0.02\, T_n$ }
\end{subfigure} 

\medskip
\medskip

\centering
\begin{subfigure}[b]{\linewidth}
\centering
\begin{subfigure}[b]{.4\linewidth}
     \includegraphics{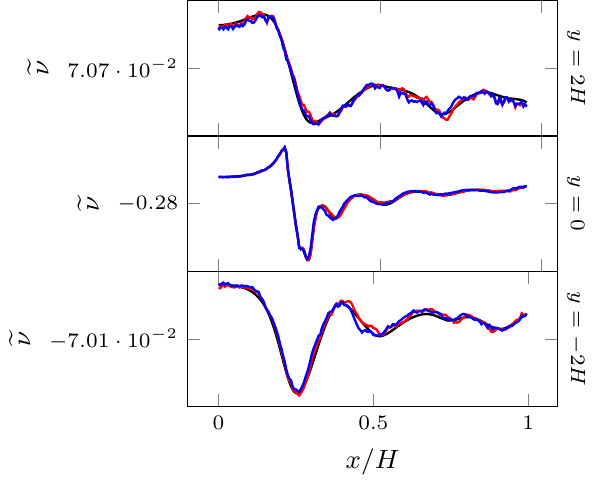}
\end{subfigure} \quad
\begin{subfigure}[b]{.4\linewidth}
      \includegraphics{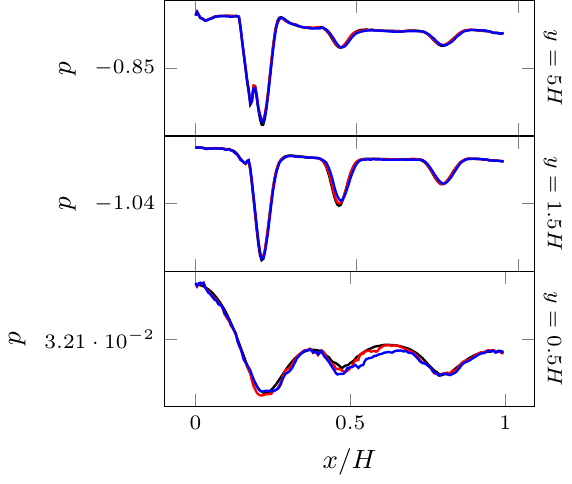}
\end{subfigure} 
\caption{Time $t= 0.33\, T_n$ }
\end{subfigure} 

\medskip
\medskip

\centering
\begin{subfigure}[b]{\linewidth}
\centering
\begin{subfigure}[b]{.4\linewidth}
      \includegraphics{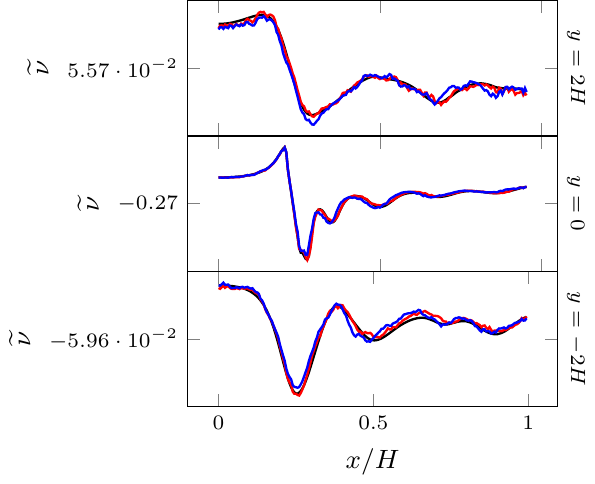}
\end{subfigure} \quad
\begin{subfigure}[b]{.4\linewidth}
         \includegraphics{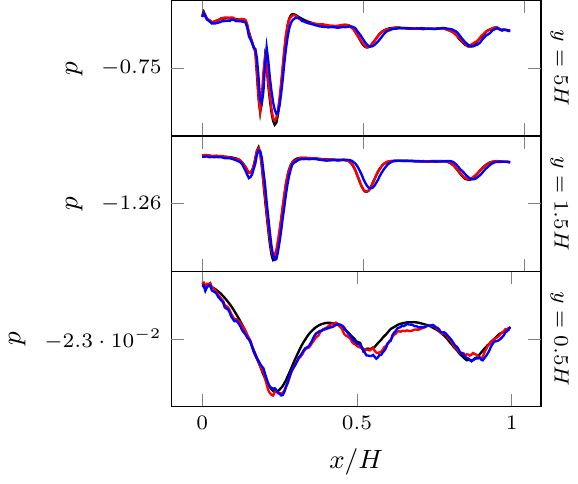}
\end{subfigure} 
\caption{Time $t= 0.66\, T_n$ }
\end{subfigure} 

\caption{\textbf{Comparative predictions of sampled along y-axis of SA turbulent viscosities for case 1 (left) and case 2 (right).} Figures from top to bottom denote the predictions at increasing times, \textit{i.e.} the top row contains instantaneous preditctions at $t= 0.02\, T_n$,  the middle row at $t= 0.33\, T_n$, and the bottom row shows the predictions at $t= 0.66\, T_n$ . }
\label{fig:comparison_pripost_TQ_aY}
\end{figure}

\begin{figure} 
\centering

\begin{subfigure}[t]{.9\linewidth}
	\centering
     \includegraphics{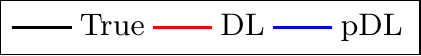}
	\caption*{}
\end{subfigure}
\vspace{-1em}%

\begin{subfigure}[b]{\linewidth}
\centering
\begin{subfigure}[b]{.4\linewidth}
     \includegraphics{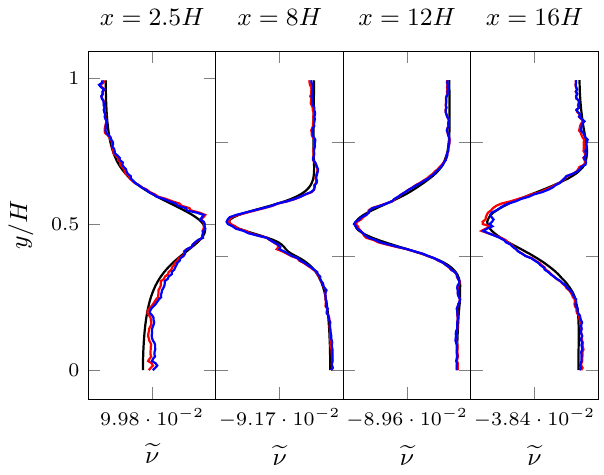}
\end{subfigure} \quad
\begin{subfigure}[b]{.4\linewidth}
      \includegraphics{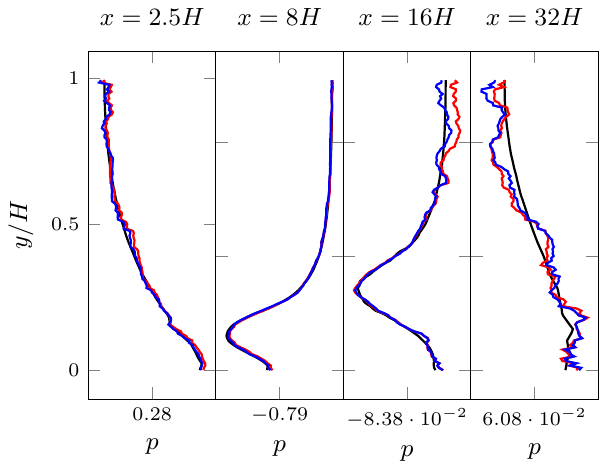}
\end{subfigure}  
\caption{Time $t= 0.02\, T_n$ }
\end{subfigure} 

\medskip
\medskip

\centering
\begin{subfigure}[b]{\linewidth}
\centering
\begin{subfigure}[b]{.4\linewidth}
     \includegraphics{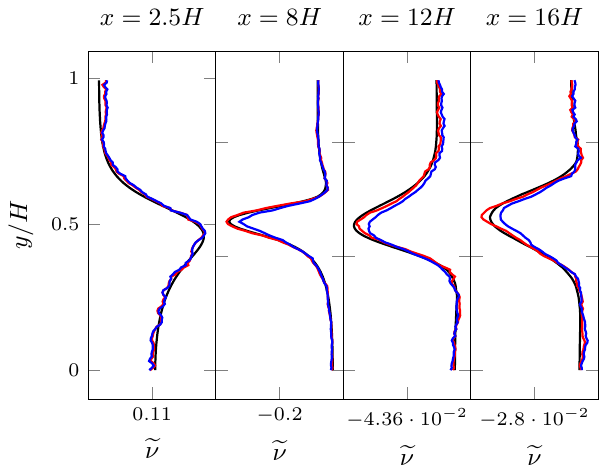}
\end{subfigure} \quad
\begin{subfigure}[b]{.4\linewidth}
      \includegraphics{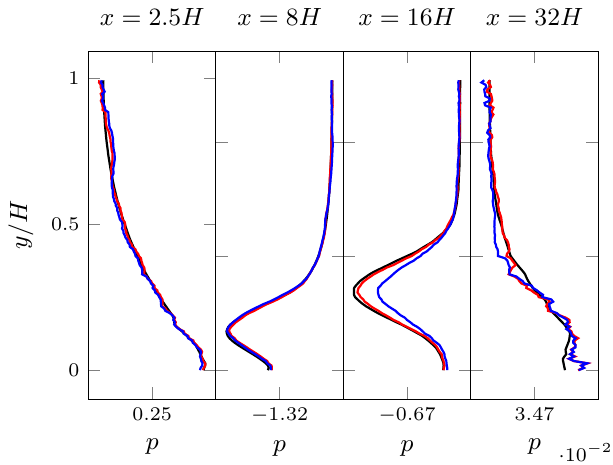}
\end{subfigure} 
\caption{Time $t= 0.33\, T_n$ }
\end{subfigure} 

\medskip
\medskip

\centering
\begin{subfigure}[b]{\linewidth}
\centering
\begin{subfigure}[b]{.4\linewidth}
      \includegraphics{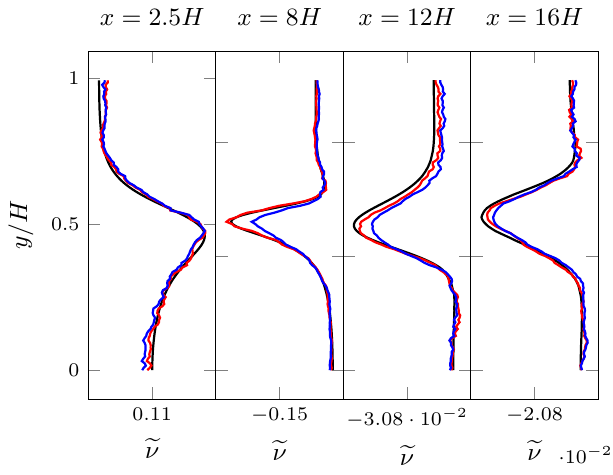}
\end{subfigure} \quad
\begin{subfigure}[b]{.4\linewidth}
         \includegraphics{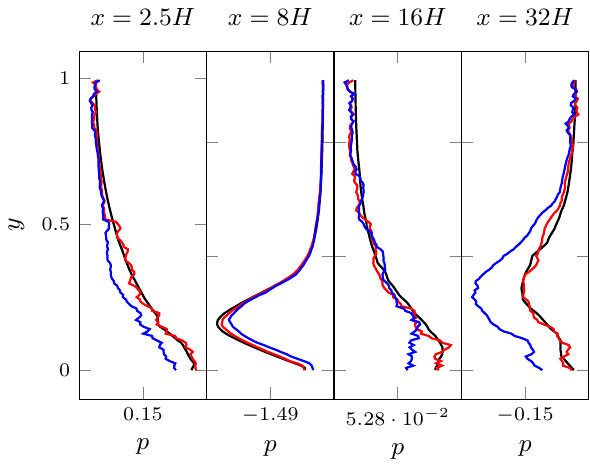}
\end{subfigure} 
\caption{Time $t= 0.66\, T_n$ }
\end{subfigure} 

\caption{\textbf{Comparative predictions of cross streamwise velocity components ($u_1$) for case 1 (left) and case 2 (right).} Figures from top to bottom, denote the predictions at increasing times, \textit{i.e.} the top row contains instantaneous predictions at $t= 0.02\, T_n$,  the middle row at $t= 0.33\, T_n$, and the bottom row shows the predictions at $t= 0.66\, T_n$.}
\label{fig:comparison_pripost_TQ}
\end{figure}

For additional verification, we show the instantaneous snapshots of cross-streamwise velocity contours for case-1 and case-2 in figure \ref{fig:cfdu1_comparison} and figure \ref{fig:cfdu1_comparisonC2} respectively.
The evolution of temporal predictions of cross streamwise velocity component when measured along the cross-streamwise directions is shown in figure \ref{fig:comparison_pripost_TV_aY}, and when measured along streamwise directions is shown in figure \ref{fig:comparison_pripost_TV}. Similarly, evolution of temporal predictions of SA turbulent viscosities when measured along the cross-streamwise directions is shown in figure \ref{fig:comparison_pripost_TQ_aY}, and when measured along streamwise directions is shown in figure \ref{fig:comparison_pripost_TQ}.

\bibliography{bibfile}

\end{document}